\documentclass[journal]{IEEEtran}

\usepackage[T1]{fontenc}
\usepackage{amsmath,amssymb,amsfonts}

\usepackage{amsthm}
\usepackage{mathtools}
\usepackage{booktabs}
\usepackage{graphicx}
\usepackage{cite}

\usepackage{enumitem}

\newtheorem{theorem}{Theorem}
\newtheorem{lemma}{Lemma}
\newtheorem{proposition}{Proposition}

\theoremstyle{definition}
\newtheorem{definition}{Definition}
\newtheorem{assumption}{Assumption}
\newtheorem{problem}{Problem}
\newtheorem{example}{Example}
\theoremstyle{remark}
\newtheorem{remark}{Remark}

\newlist{condC}{enumerate}{1}
\setlist[condC]{label=\textup{(C\arabic*)},ref=\textup{(C\arabic*)},leftmargin=*,itemsep=1pt,topsep=2pt}
\newlist{condD}{enumerate}{1}
\setlist[condD]{label=\textup{(D\arabic*)},ref=\textup{(D\arabic*)},leftmargin=*,itemsep=1pt,topsep=2pt}
\newlist{condF}{enumerate}{1}
\setlist[condF]{label=\textup{(F\arabic*)},ref=\textup{(F\arabic*)},leftmargin=*,itemsep=1pt,topsep=2pt}
\newlist{condR}{enumerate}{1}
\setlist[condR]{label=\textup{(R\arabic*)},ref=\textup{(R\arabic*)},leftmargin=*,itemsep=1pt,topsep=2pt}
\newlist{condL}{enumerate}{1}
\setlist[condL]{label=\textup{(L\arabic*)},ref=\textup{(L\arabic*)},leftmargin=*,itemsep=1pt,topsep=2pt}
\newlist{romanlist}{enumerate}{1}
\setlist[romanlist]{label=\textup{(\roman*)},ref=\textup{(\roman*)},leftmargin=*,itemsep=1pt,topsep=2pt}

\newcommand{\E}{\mathbb{E}}
\newcommand{\Prob}{\mathbb{P}}
\newcommand{\R}{\mathbb{R}}
\newcommand{\N}{\mathbb{N}}
\newcommand{\States}{S}
\newcommand{\hitz}{\sigma_z}
\newcommand{\retz}{\tau_z}
\newcommand{\Jbar}{\overline{J}}
\newcommand{\taboo}{\widetilde{P}}
\newcommand{\Inv}{\mathsf{Inv}}
\newcommand{\sr}{\operatorname{sr}}
\newcommand{\tr}{\operatorname{tr}}
\newcommand{\ind}{\mathbf{1}}
\allowdisplaybreaks

\usepackage{tikz}
\newcommand{\qdx}{1.25}
\tikzset{>=stealth,
  qn/.style={circle,draw=black!70,fill=black!10,thick,inner sep=1pt,minimum size=16pt,font=\footnotesize},
  qref/.style={qn,double,double distance=1.2pt},
  qabs/.style={qn,fill=black!30},
  qsub/.style={fill=black!35},
  ev/.style={font=\scriptsize},
  del/.style={->,dashed,draw=black!45}}
\newcommand{\queuenodes}{%
  \node[qref] (n0) at (0,0) {$0$};
  \node[qn] (n1) at (\qdx,0) {$1$};
  \node[qn] (n2) at (2*\qdx,0) {$2$};
  \node[qn] (n3) at (3*\qdx,0) {$3$};
  \node[qn] (n4) at (4*\qdx,0) {$4$};
  \node (nd) at (4.75*\qdx,0) {$\cdots$};}
\newcommand{\queueedges}[2]{%
  \foreach \a/\b in {n0/n1,n1/n2,n2/n3,n3/n4,n4/nd} \draw[->] (\a) to[bend left=40] node[ev,above]{#1} (\b);
  \foreach \a/\b in {n2/n1,n3/n2,n4/n3,nd/n4} \draw[->] (\a) to[bend left=40] node[ev,below]{#2} (\b);}

\begin{document}

\title{Complete Certificates for Long-Run Average Costs}

\author{ Arash Bahari Kordabad, Rupak Majumdar, V. R.  Sathiyanarayana, Sadegh Soudjani%
\thanks{All authors are with the Max Planck Institute for Software Systems, Kaiserslautern, Germany. S. Soudjani is also with the School of Computer Science, University of Birmingham, Birmingham, United Kingdom. Corresponding author: Arash Bahari Kordabad (e-mail: \mbox{arashbk@mpi-sws.org}).}%
}

\maketitle

\begin{abstract}
This paper provides certificates for verifying
lower and upper bounds on the long-run average cost of Markov chains on countable state spaces.
Our results generalize existing results on finite state spaces,
where a bound is certified by a bias function that satisfies a one-step inequality at every state.
On a countably infinite state space, this inequality is not sufficient and classical verification theorems add a growth condition on the bias. 
This condition is a limit along the trajectories of the chain 
and cannot be locally checked at each transition.
We replace it with locally checkable conditions in two steps.
We first bound the bias by a function that is superharmonic for the chain killed at a reference state and whose harmonic part is zero at that state.
The superharmonicity of this function is locally checkable, but the condition on its harmonic part is not and involves a limit.
We then show that the harmonic part is zero exactly when an auxiliary chain built from that function is absorbed almost surely.
Finally, we certify the absorption using a sound and complete locally checkable certificate for almost-sure reachability.
We show that the rule is sound for every countable chain and complete for irreducible positive recurrent chains whose expected absolute cost in steady state is finite.
We apply the rule to a networked control system with a packet-dropping channel, where the certificate is computed from two Lyapunov equations and recovers the critical packet arrival probability.
\end{abstract}


\section{Introduction}
\label{sec:introduction}

\IEEEPARstart{T}{he} long-run average cost is a commonly used performance measure of a stochastic system that runs without a terminal time.
In networked control, it represents the mean estimation cost per step of a loop that receives its measurements over a lossy channel~\cite{YatesEtAl2021,ChampatiEtAl2019,KluegelEtAl2019,LiuEtAl2022}.
In Markov decision theory, it is the criterion of the average-cost problem~\cite{Puterman94,ABFGM93}.
For a Markov chain with stationary distribution $\pi$ and cost $c$, this long-run average cost is $\pi(c)$ under some model assumptions.
The stationary distribution is known in closed form only for chains with special structure.
On an infinite state space, there is no general method to compute it.

We consider the problem of verifying that the long-run average cost of 
a system modeled as an infinite-state Markov chain lies between given bounds.
We take the approach of locally checkable certificates: given a system and potential upper and lower bounds, we reduce the problem to finding a finite set of certificate
functions that satisfy inequalities
over the certificates and one transition of the chain.
Our main result provides locally checkable certificates for the long-run average cost of countable-state
Markov chains.
We show that the certificates are \emph{sound}: they are sufficient conditions for every countable Markov
chain.
We also show a partial \emph{completeness} (or converse) theorem: 
our certificates exist whenever the bound is valid, 
provided the chain is irreducible positive recurrent and its expected absolute cost in steady state
is finite.

For the problem of almost-sure reachability,
sound and complete locally checkable certificates exist.
There, Foster's criterion certifies positive recurrence by a one-step drift inequality~\cite{Foster53}, ranking supermartingales certify almost-sure termination of probabilistic programs~\cite{ChakarovS13}, and the martingale rule of~\cite{MS25} and the rule of~\cite{McIverMKK18} provide locally checkable certificates for almost-sure reachability of countable Markov chains that are both sound and complete.
All conditions of these rules are inequalities over one transition.
For expected values the same is known to fail for lower bounds, which need a side condition~\cite{HKKM20}.

We generalize these techniques substantially and provide, to the best of our knowledge, 
the first sound and complete locally checkable
certificates for long-run average costs.

\smallskip
\emph{Our approach.}
Our starting point uses an existing characterization for long-run average costs
\cite{Puterman94,Derman66,ABFGM93}.
The characterization uses a scalar $\rho$ and a \emph{bias function} $h$ on the state space 
that satisfies the Poisson inequality $\rho+h\le c+Ph$ at every state, where $P$ is the transition kernel.
On a finite state space this inequality implies that the long-run average cost is at least $\rho$, and the reversed inequality implies that 
it is at most $\rho$~\cite[Thm.~8.4.1]{Puterman94}.
However, on a countable state space, this implication can fail for an unbounded bias, even when the cost is bounded: as we demonstrate in Example~\ref{ex:false}
later, there are positive recurrent chains with bounds above the true value that satisfy the inequality at every state.
Proofs of average-cost bounds therefore add conditions constraining the growth of the bias.
In~\cite{Derman66}, the bias is required to be bounded from above and from below, and 
in \cite[Thm.~5.1]{ABFGM93}, it is required to grow sublinearly in expectation along the chain.
This sublinear-growth condition involves a limit over the distribution of the chain at large times, and cannot be checked locally by an inequality at each state.

We start from the Poisson inequality together with the growth condition of \cite[Thm.~5.1]{ABFGM93}, and we replace the growth condition by a Lyapunov certificate.
Lyapunov conditions make the growth condition locally checkable and have been proposed before.
For example, a geometric drift bounding the growth is a one-step inequality, but it holds only on a subclass of the positive recurrent chains~\cite{MeynTweedie93}.
A similar Lyapunov condition in~\cite[Thm.~5.1]{Hordijk74}, \cite[Assumption~5.13]{ABFGM93} covers a larger class, but it contains a decay clause that is again a limit.
In both cases, the condition is stated as a sufficient one, and no converse is available that states that a certificate exists whenever the bound is valid.

We proceed differently to these works and obtain both soundness and completeness.
We show that the bias is dominated by a function that is superharmonic for the chain killed at a reference state and whose harmonic part is zero at that state.
We prove that this rule is sound for every countable chain and complete for irreducible positive recurrent chains whose expected absolute cost in steady state is finite.
A single bias, built from the first-passage costs to the reference state, certifies every valid lower bound and every valid upper bound.
This completeness is not contained in~\cite{ABFGM93}, where the Lyapunov condition is only shown to be sufficient.

Next, we prove that the harmonic part is zero if and only if an auxiliary chain, obtained from the killed chain by Doob's transform \cite{Doob84} with the Lyapunov function, is absorbed almost surely.
We certify the absorption by the martingale rule of~\cite{MS25}, which is sound and complete.
Altogether, this gives the \emph{local rule}, which is our main result: 
a locally checkable sound and complete proof rule for the long-run average cost with no condition on trajectories or limits along the chain.

As an application, we compute a local certificate for a networked control system whose controller receives the state over a packet-dropping channel.
For a plant with spectral radius greater than one, the cost grows exponentially with the age of the last received state.
The certificate follows from two Lyapunov equations and holds on the whole countable state space.
A certificate for a finite upper bound exists exactly when the packet arrival probability exceeds the known critical value~\cite{SinopoliEtAl2004}.

\emph{Related work.}
Our verification theorem and its proof specialize theorems for the average-cost optimality equation to chains without control.
The growth condition appears for countable controlled chains in \cite[Thm.~5.1]{ABFGM93}, and goes back to~\cite{Derman66}.
The one-sided forms of the rule correspond to the average-cost optimality inequality~\cite{Sennott89}.
For finite chains the gain and the bias are given by the limiting matrix and the deviation matrix of the chain~\cite{Puterman94,LamondPuterman89}.
The Poisson equation on general state spaces is treated in \cite[Sec.~17.4]{MeynTweedie93}.
The bias that we use for completeness is the first-passage solution of~\cite[Thm.~1]{DermanVeinott67}, which is also the cost potential of Markov potential theory~\cite[Sec.~5]{Hordijk74}.

Recurrence hypotheses are needed for completeness.
This is consistent with the counterexample in~\cite{CavazosCadena91}, in which the average-cost optimality equation has no solution in the absence of positive recurrence.
Our transformed chain is the Doob $h$-transform~\cite{Doob84} of the killed chain, with the Lyapunov function in the role of $h$, and potentials and harmonic functions of countable chains are developed in~\cite{KemenySnellKnapp76}.
The characterization of recurrence by inequalities in~\cite{MSZ78} is the basis of the completeness proof of the reachability rule that we use.

Lyapunov functions and barrier certificates verify stability and safety of infinite-state systems by inequalities that are checked at each state~\cite{PrajnaJadbabaiePappas07}.
Converse theorems state when such a certificate is also necessary.
It is shown in~\cite{LinSontagWang96} that every robustly globally asymptotically stable system admits a smooth Lyapunov function, and in~\cite{Ratschan18} that every robustly safe ordinary differential equation admits a barrier certificate.
For discrete-time stochastic systems, a converse Lyapunov theorem for asymptotic stability in probability is proved in~\cite{TeelHespanhaSubbaraman14}, and barrier-like conditions that are necessary and sufficient for safety and reach-avoid verification are derived in~\cite{Xue26}.
It is shown in~\cite{MSS24} that drift and variant conditions are necessary and sufficient for almost-sure reachability.
These certificates are computed by sum-of-squares programs for polynomial dynamics in~\cite{KMS26}, and their existence is characterized for linear systems in~\cite{KMMS26}.
Necessary and sufficient conditions are also given for set invariance with probability one of controlled diffusions in~\cite{WangEtAl25}, and a synthesis procedure for supermartingale certificates of $\omega$-regular specifications that is sound and complete for polynomial templates is given in~\cite{AbateGiacobbeRoy24}.
The local rule provides such a necessary and sufficient certificate for the long-run average cost.

\emph{Outline.}
The rest of the paper is organized as follows.
Section~\ref{sec:setting} introduces the model and the problem.
Section~\ref{sec:rule} states the verification theorem with its growth condition and constructs the canonical bias.
Section~\ref{sec:fl} gives the Lyapunov rule and Section~\ref{sec:local} the local rule.
Section~\ref{sec:networked_control} presents the case study, and Section~\ref{sec:conclusion} concludes the paper.

\section{Setting and Problem Statement}
\label{sec:setting}

\subsection{Markov Chains with Long-Run Average Cost}
\label{sec:chains}

We write $\N=\{0,1,2,\dots\}$ and $\R$ for the real numbers.
For an event $A$, we write $\ind_A$ for its indicator,
equal to $1$ if $A$ occurs and $0$ otherwise. Let $(X_t)_{t\in\N}$ be a Markov chain on a countable
state space $\States$ with time-independent transition
probabilities $p(y\mid x)$, $x,y\in\States$
(i.e., a time-homogeneous Markov chain).
We denote its transition kernel by
$P:\States\times\States\to[0,1]$, defined by
$P(x,y)=p(y\mid x)$, with
$\sum_{y\in\States}P(x,y)=1$ for every $x\in\States$.
A possibly unbounded cost function $c:\States\to\R$
assigns a cost $c(x)$ to each state $x\in\States$.
We write $\E_x$ and $\Prob_x$ for expectation and probability
when the chain starts at $X_0=x$, and
$\mathcal F_t=\sigma(X_0,\dots,X_t)$ for the $\sigma$-algebra
representing the information contained in the chain's history
up to time $t$. For a function $h:\States\to\R$, we write
$h^+:=\max\{h,0\}$ and $h^-:=\max\{-h,0\}$.
We set $(Ph)(x)=\sum_{y\in\States}P(x,y)h(y)$
whenever the sum is well defined, i.e., at least one of
$\sum_{y\in\States}P(x,y)h^+(y)$ and
$\sum_{y\in\States}P(x,y)h^-(y)$ is finite.
For a measure $\mu$ on $\States$, assigning a finite
nonnegative weight $\mu(x):=\mu(\{x\})$ to each state $x$,
we set $\mu(h)=\sum_{x\in\States}\mu(x)h(x)$
whenever the sum is well defined. A function $Q:\States\times\States\to[0,\infty)$
is called a kernel.
It is stochastic if $\sum_{y\in\States}Q(x,y)=1$
for every $x\in\States$, and substochastic if
$\sum_{y\in\States}Q(x,y)\le1$ for every $x\in\States$.
Thus, $P$ is stochastic, and every stochastic kernel
is substochastic.
For a function $f:\States\to\R$, we set
$(Qf)(x)=\sum_{y\in\States}Q(x,y)f(y)$
whenever the sum is well defined, in the same sense as above. 
For a substochastic kernel $Q$, we define
$Q^0(x,y)=1$ if $x=y$ and $Q^0(x,y)=0$ otherwise.
For each $n\ge1$, we recursively define
$(Q^n)(x,y)=\sum_{u\in\States}(Q^{n-1})(x,u)Q(u,y)$,
$x,y\in\States$.

\begin{definition}[Long-run average cost]
\label{def:avg}
The lower and upper expected long-run average costs
from the initial state $x\in\States$ are defined by
\begin{align*}
  J(x)&=\liminf_{N\to\infty}\frac1N\,\E_x\Bigl[\sum_{t=0}^{N-1}c(X_t)\Bigr],\\
  \Jbar(x)&=\limsup_{N\to\infty}\frac1N\,\E_x\Bigl[\sum_{t=0}^{N-1}c(X_t)\Bigr],
\end{align*}
provided that the expectations are well defined for every
$N\ge1$. These values may be infinite, and $J(x)\le\Jbar(x)$.
\end{definition} 
The expected average cost need not converge as the horizon
$N$ increases. For example, let $\States=\N$ and
$p(x+1\mid x)=1$ for every $x\in\States$, so that the
state increases by one at each step.
Set $c(0)=0$, $c(1)=c(2)=1$, $c(3)=\cdots=c(6)=0$,
and $c(7)=\cdots=c(14)=1$, continuing with alternating
blocks of zeros and ones whose lengths double.
At the ends of the zero blocks, the average cost approaches
$1/3$, whereas at the ends of the one blocks, it approaches
$2/3$. Thus, the average cost oscillates between these
two values in the long run and has no limit.
In this example, $J(x)=1/3$ and $\Jbar(x)=2/3$
for every initial state $x\in\States$.
We therefore state lower bounds for $J$ and upper bounds
for $\Jbar$.
\begin{problem}
\label{prob:main}
Let the chain $P$, defined on the countable
state space $\States$, the cost $c$ and a scalar $\rho\in\R$ be given.
Find conditions on finitely many functions on $\States$ such that
\begin{romanlist}
\item every condition is an inequality that involves the values of the functions at a state $x$, the transition probabilities $p(\cdot\mid x)$, and the values of the functions at the successors of $x$,
\item the conditions imply $J(x)\ge\rho$ for every $x\in\States$,
\item the functions exist whenever the bound $J\ge\rho$ is valid.
\end{romanlist}
We seek analogous conditions for the upper bound $\Jbar(x)\le\rho$.
\end{problem}

Property (ii) is soundness and property (iii) is completeness.
We prove soundness for every countable chain and completeness on the class of chains defined in the next subsection.

\subsection{Irreducible Positive Recurrent Chains}
\label{sec:model}

A state $y$ is accessible from $x$ if $P^n(x,y)>0$ for some $n\in\N$.
The chain is irreducible if every state is accessible from every other state.
For a state $x$ let $\sigma_x=\inf\{t\ge0 \mid X_t=x\}$ and $\tau_x=\inf\{t\ge1 \mid X_t=x\}$, with $\inf\emptyset=\infty$.
The chain is positive recurrent if $\E_x[\tau_x]<\infty$ for every $x$.
A probability measure $\pi$ on $\States$ is a stationary distribution if $\sum_x\pi(x)p(y\mid x)=\pi(y)$ for every $y$. Every irreducible positive recurrent Markov chain on a countable state space admits a unique stationary distribution \cite[Ch.~1]{Norris97}.

For an irreducible chain,
Foster's criterion provides a sufficient condition for
positive recurrence.
Suppose there exist a Foster--Lyapunov function
$v:\States\to[0,\infty)$,
a nonempty finite set $\mathcal C\subseteq\States$, and a constant
$\delta>0$ such that $(Pv)(x)-v(x)\le-\delta$  for every $x\in\States\setminus \mathcal C,$ and $(Pv)(x)<\infty$ for every $x\in \mathcal C$.
Then the chain is positive recurrent
\cite[Thm.~11.3.4]{MeynTweedie93}.

\begin{assumption}[Irreducible Positive Recurrent Chains]
\label{ass:model}
The chain is irreducible and positive recurrent with stationary distribution $\pi$, and the cost satisfies $\pi(|c|)<\infty$.
We set $\rho^*=\pi(c)$.
\end{assumption}

\begin{example}[A single-server queue]
\label{ex:queue}
Consider a queue with a single server, and let the state $x\in\States=\N$ be the number of customers in the queue.
In each step a new customer arrives with probability $\alpha\in(0,\tfrac12)$, and otherwise, with probability $\beta=1-\alpha$, the server completes the service of one customer if the queue is not empty.
Hence, from $x\ge1$ the chain moves to $x+1$ with probability $\alpha$ and to $x-1$ with probability $\beta$, and from $0$ it moves to $1$ with probability $\alpha$ and stays at $0$ with probability $\beta$.
The cost is $c=\ind_{\{0\}}$. The chain is irreducible because $\alpha,\beta>0$,
so any state can be reached from any other state through
a finite sequence of arrivals or service completions.
To apply Foster's criterion, choose $v(x)=x$, $\mathcal C=\{0\}$, and
$\delta=\beta-\alpha>0$.
For every $x\ge1$,
$(Pv)(x)-v(x)=\alpha-\beta=-\delta$,
while $(Pv)(0)=\alpha<\infty$.
Thus, the chain is also positive recurrent and it admits a stationary distribution. To compute the stationary distribution, we solve the
balance equations
$\pi(x)\alpha=\pi(x+1)\beta$, $x\in\N$,
which give $\pi(x)=(\alpha/\beta)^x\pi(0)$.
Since $\alpha<\beta$, requiring the probabilities
to sum to one gives
$1=\sum_{x=0}^{\infty}\pi(x)
=\pi(0)/(1-\alpha/\beta)$.
Thus, $\pi(x)=(1-\alpha/\beta)(\alpha/\beta)^x$
is a stationary distribution.
Moreover, $0\le c\le1$, so $\pi(|c|)\le1<\infty$.
Therefore, Assumption~\ref{ass:model} holds, with
$\rho^*=\pi(c)=\pi(0)=(\beta-\alpha)/\beta$.
\end{example}

To express the long-run average cost using the cost and
duration between successive visits to a state, we fix
a reference state $z\in\States$. Under Assumption~\ref{ass:model}, the cycle formulas of \cite[Sec.~1.7]{Norris97} state that
\begin{equation}
\label{eq:kac}
  \E_z[\retz]=\frac{1}{\pi(z)},\qquad
  \E_z\Bigl[\sum_{t=0}^{\retz-1}f(X_t)\Bigr]=\frac{\pi(f)}{\pi(z)}
\end{equation}
for every $f:\States\to[0,\infty)$.
The chain visits $z$ once in $1/\pi(z)$ steps on average, and the expected total of $f$ between two visits is the 
expected value of $f$ in steady state divided by $\pi(z)$.

The next two lemmas ensure that our certificate functions
are finite and identify the long-run average cost,
respectively \cite[Ch.~14]{MeynTweedie93},
\cite[Ch.~VI]{Asmussen03}.

\begin{lemma}[Finite first-passage totals]
\label{lem:hitting}
Let Assumption~\ref{ass:model} hold and let $f:\States\to[0,\infty)$ satisfy $\pi(f)<\infty$.
Then $\E_x\bigl[\sum_{t=0}^{\hitz-1}f(X_t)\bigr]<\infty$ for every $x\in\States$.
\end{lemma}

\begin{lemma}[Convergence of the averages]
\label{lem:cesaro}
Let Assumption~\ref{ass:model} hold.
Then
\begin{equation*}
  \lim_{N\to\infty}\frac1N\,\E_x\Bigl[\sum_{t=0}^{N-1}c(X_t)\Bigr]=\rho^*
\end{equation*}
for every $x\in\States$.
Hence $J(x)=\Jbar(x)=\rho^*$ for every $x$. We call $\rho^*$ the \emph{gain}. 
\end{lemma}

Under Assumption~\ref{ass:model} the long-run average cost does not depend on the initial state.
A lower (upper) bound $\rho$ is valid if and only if $\rho\le\rho^*$ ($\rho\ge\rho^*$).
We also use the recentered cost and its first-passage total
\begin{equation}
\label{eq:G}
  c'(x):=|c(x)-\rho^*|,\qquad
  G(x):=\E_x\Bigl[\sum_{t=0}^{\hitz-1}c'(X_t)\Bigr].
\end{equation}
By the triangle inequality, $
c'(x)\le |c(x)|+|\rho^*|$.
Since $\rho^*=\pi(c)$, we have
$
|\rho^*|=|\pi(c)|\le \pi(|c|)<\infty,
$
where the last inequality follows from Assumption~\ref{ass:model}. Since $\pi$ is a probability measure, $
\pi(c')
\le \pi(|c|)+|\rho^*|
<\infty.
$
Thus, $c'$ satisfies the hypotheses of Lemma~\ref{lem:hitting}, and applying the lemma with $f=c'$ gives
$
G(x)
=
\E_x\Bigl[\sum_{t=0}^{\hitz-1}c'(X_t)\Bigr]
<\infty$ 
for every $x\in\States$.
Applying the same lemma with the constant function $f=1$, for which
$\pi(1)=1<\infty$, gives
$
\E_x[\hitz]
=
\E_x\Bigl[\sum_{t=0}^{\hitz-1}1\Bigr]
<\infty$ for every $x\in\States$.

\subsection{The Killed Chain}
\label{sec:killed}

\begin{definition}[Taboo kernel]
\label{def:taboo}
The taboo kernel at the reference state $z$ is
\begin{equation*}
  (\taboo f)(x)=\sum_{y\ne z}p(y\mid x)f(y).
\end{equation*}
\end{definition}

The taboo kernel removes the probability mass of transitions into $z$, and therefore it is substochastic and loses the mass $p(z\mid x)$ at $x$.
Thus, the chain with kernel $\taboo$ can be viewed as the original
chain killed upon its next visit to $z$, and we call it the
\emph{killed chain}.

For every $f:\States\to[0,\infty)$, induction on $n$ gives
\begin{equation}
\label{eq:taboo-iter}
  (\taboo^{\,n}f)(x)=\E_x\bigl[f(X_n)\ind_{\{\retz>n\}}\bigr].
\end{equation}
Thus, $\taboo^{\,n}f$ computes exactly the contribution of trajectories
that have not visited $z$ during the times $1,\ldots,n$.
Since $f\ge0$, Tonelli's theorem \cite[Sec.~18]{Billingsley95}
allows the sum over $n$ and the expectation to be interchanged:
\begin{align}
  \sum_{n=0}^{\infty}(\taboo^{\,n}f)(x)
  &=
  \sum_{n=0}^{\infty}
  \E_x\bigl[f(X_n)\ind_{\{\retz>n\}}\bigr] \notag\\
  &=
  \E_x\Bigl[
    \sum_{n=0}^{\infty}
    f(X_n)\ind_{\{\retz>n\}}
  \Bigr] \notag\\
  &=
  \E_x\Bigl[
    \sum_{n=0}^{\retz-1}f(X_n)
  \Bigr].
\label{eq:expected-total}
\end{align}
Hence, $\sum_{n\ge0}\taboo^{\,n}f$ is the expected accumulated
$f$-cost up to the next visit to $z$.

A function $w:\States\to[0,\infty)$ is \emph{superharmonic} for $\taboo$ if $\taboo w\le w$ pointwise, and \emph{harmonic} if $\taboo w=w$.
Since $\taboo$ is a nonnegative kernel, it preserves
pointwise inequalities: $f\le g$ implies
$\taboo f\le\taboo g$. Therefore a superharmonic $w$ satisfies $0 \leq \taboo^{\,n+1}w\le\taboo^{\,n}w\le w$ for every $n$ and the pointwise limit $w_\infty=\lim_{n\to\infty}\taboo^{\,n}w$ exists.
This limit $w_\infty$ is harmonic, and $w$ has the decomposition~\cite{KemenySnellKnapp76}
\begin{equation}
\label{eq:riesz}
  w=
  \sum_{n=0}^{\infty}
  \taboo^{\,n}\bigl(w-\taboo w\bigr)
  +w_\infty.
\end{equation}
By~\eqref{eq:expected-total}, the first term is the expected
accumulated one-step decrement $w-\taboo w$ up to the next visit to
$z$.
The derivation of~\eqref{eq:riesz} and of the harmonicity of $w_\infty$ is presented in Appendix~\ref{app:riesz}.
We call $w_\infty$ the \emph{harmonic part} of $w$.
A harmonic component cancels out in the one-step decrements, i.e., if $e\ge0$ is harmonic, then $(w+e)-\taboo(w+e)=w-\taboo w$, and $w$ and $w+e$ have the same one-step decrements.

We have now established the probabilistic objects needed for the certificate rules presented in the next sections.

\section{The Poisson Inequality}
\label{sec:rule}

We begin with the classical verification theorem.
A scalar $\rho$ and a bias function $h$ that satisfy the one-step Poisson inequality at every state certify a lower bound $\rho$, provided that the bias satisfies a growth condition along the chain.
On a finite state space the growth condition holds for every bias, and the Poisson inequality alone is sound and complete \cite[Ch.~8]{Puterman94}.
On a countable state space the growth condition is needed, and it is an asymptotic condition rather than a one-step inequality.
This section states this verification theorem,
constructs the canonical bias that the completeness proofs use, and shows by an example that the growth condition cannot be omitted.

\subsection{Certificates with a Growth Condition}

\begin{definition}[Certificates with a Growth Condition]
\label{def:cert}
A pair $(\rho,h)$ with $\rho\in\R$ and $h:\States\to\R$ is a \emph{lower-bound certificate} if
\begin{condC}
\item \label{itm:c1} for every $x\in\States$, $(Ph)(x)$ is well defined in $\R$ and
\begin{equation*}
  \rho+h(x)\le c(x)+(Ph)(x),
\end{equation*}
\item \label{itm:c2} $\E_x[|h(X_t)|]<\infty$ for every $x\in\States$ and $t\in\N$, and
\begin{equation*}
  \limsup_{N\to\infty}\frac1N\,\E_x[h(X_N)]\le0\quad\text{for every }x\in\States.
\end{equation*}
\end{condC}
It is an \emph{upper-bound certificate} if
\begin{condD}
\item \label{itm:d1} for every $x\in\States$, $(Ph)(x)$ is well defined in $\R$ and
\begin{equation*}
  \rho+h(x)\ge c(x)+(Ph)(x),
\end{equation*}
\item \label{itm:d2} $\E_x[|h(X_t)|]<\infty$ for every $x\in\States$ and $t\in\N$, and
\begin{equation*}
  \liminf_{N\to\infty}\frac1N\,\E_x[h(X_N)]\ge0\quad\text{for every }x\in\States.
\end{equation*}
\end{condD}
\end{definition}
We call~\ref{itm:c1} the \emph{lower Poisson inequality} and~\ref{itm:d1} the \emph{upper Poisson inequality}, after the Poisson inequality of \cite[App.~A]{Meyn08}, and we call~\ref{itm:c2} and~\ref{itm:d2} the \emph{growth condition}.
The scalar $\rho$ is the claimed gain, while $h$ is a bias function.
Since $P1=1$, adding a constant to $h$ changes neither $Ph-h$ nor the limits in the growth condition.
Condition~\ref{itm:c1} can be written as
\begin{equation*}
  \rho
  \le
  c(x)+(Ph)(x)-h(x)
  =
  c(x)+
  \E_x\!\left[h(X_1)-h(X_0)\right].
\end{equation*}
Thus, at every state, the sum of the cost and the expected one-step change in the bias is at least $\rho$.
The reversed inequality in~\ref{itm:d1} has the analogous interpretation for an upper bound.
The one-step inequalities (C1) and (D1) alone are insufficient on an infinite state space.
Summing them over a finite horizon produces the boundary term $\E_x[h(X_N)]$, and the growth condition controls this term as $N\to\infty$.

Following~\cite{Mandl74}, define \emph{slack} to be
$$
  \Phi(x)=c(x)+(Ph)(x)-\rho-h(x).
$$
Thus, the lower Poisson inequality is equivalent to a nonnegative slack, the upper one to
a nonpositive slack, and $\Phi=0$ is the Poisson equation
$\rho+h=c+Ph$ with zero slack.

\begin{theorem}[Verification theorem]
\label{thm:expected}
Let $(\rho,h)$ be a lower-bound certificate.
Then $\E_x[c(X_t)^-]<\infty$ for every $x$ and $t$, the expectations in Definition~\ref{def:avg} are well defined in $(-\infty,\infty]$, and $J(x)\ge\rho$ for every $x\in\States$.

Similarly, let $(\rho,h)$ be an upper-bound certificate.
Then $\E_x[c(X_t)^+]<\infty$ for every $x$ and $t$, the expectations are well defined in $[-\infty,\infty)$, and $\Jbar(x)\le\rho$ for every $x\in\States$.
\end{theorem}

The proof is given in Appendix~\ref{app:expected}.
For a lower-bound certificate, it shows that
\begin{equation}
\label{eq:telescope}
  N\rho+h(x)\le\E_x\Bigl[\sum_{t=0}^{N-1}c(X_t)\Bigr]+\E_x[h(X_N)]
\end{equation}
for every $N\ge1$.

Theorem~\ref{thm:expected} imposes no additional assumptions
on the system model or the cost.
The integrability clause of the growth condition ensures that the expectations involving the bias are well defined and justifies the conditional-expectation and cancellation steps that lead to Eq.~\eqref{eq:telescope}.

\begin{example}[Certificate for the single-server queue]
\label{ex:cert}
For the queue of Example~\ref{ex:queue}, choose any
constant $\theta\in\R$, let
$h(x)=\theta-x/\beta$ for every $x\in\N$, and fix
$\rho\le\rho^*$. We first check the lower Poisson inequality.
For $x\ge1$, the cost is zero, and
\begin{equation*}
  c(x)+(Ph)(x)-h(x)
  =
  \frac{\beta-\alpha}{\beta}
  =
  \rho^*.
\end{equation*}
At $x=0$, since $(Ph)(0)=\alpha h(1)+\beta h(0)$, we obtain
\begin{equation*}
  c(0)+(Ph)(0)-h(0)
  =
  1+\alpha\bigl(h(1)-h(0)\bigr)
  =
  1-\frac{\alpha}{\beta}
  =
  \rho^*.
\end{equation*}
Thus, condition~\ref{itm:c1} holds at every state,
with slack $\Phi=\rho^*-\rho\ge0$.

We next check the growth condition~\ref{itm:c2}.
Since the queue length can increase by at most one
per step, $X_t\le x+t$ almost surely when $X_0=x$. Hence,
\[
  \E_x[|h(X_t)|]
  \le
  |\theta|+\frac{\E_x[X_t]}{\beta}
  \le
  |\theta|+ \frac{x+t}{\beta}
  <\infty
\]
for every $x\in\N$ and $t\in\N$.
Moreover, $h\le\theta$, so
\[
  \limsup_{N\to\infty}
  \frac{1}{N}\E_x[h(X_N)]\le
  \lim_{N\to\infty}\frac{\theta}{N}
  =
  0.
\]
Therefore, $(\rho,h)$ is a lower-bound certificate
for every $\rho\le\rho^*$ and $\theta\in\R$.
By Theorem~\ref{thm:expected}, $J(x)\ge\rho$
for every $x\in\N$.
In particular, choosing $\rho=\rho^*$ gives
$J(x)\ge\rho^*$ for every $x\in\N$.
\end{example}

\subsection{The Canonical Bias}
\label{sec:completeness}

The completeness proofs of the Lyapunov rule and the local rule in the next two sections use a single bias, which certifies every valid lower bound and every valid upper bound.
Under Assumption~\ref{ass:model}, we construct this bias from the cost accumulated before the first visit to the reference state $z$, centered at the gain $\rho^*$.
This first-passage representation of the bias is classical.
It is the solution of the countable-state Poisson equation given in~\cite[Thm.~1]{DermanVeinott67} through the expected time and the expected cost of a first passage to a fixed state, and it is the potential of the cost centered at the gain for the taboo kernel, which appears for controlled chains in~\cite[Sec.~5]{Hordijk74}.

\begin{definition}[Canonical bias]
\label{def:canonical}
Under Assumption~\ref{ass:model} the canonical bias is
\begin{equation*}
  h^*(x)=\E_x\Bigl[\sum_{t=0}^{\hitz-1}\bigl(c(X_t)-\rho^*\bigr)\Bigr],\qquad x\in\States.
\end{equation*}
\end{definition}
Thus, $h^*(x)$ is the expected excess cost accumulated before the
first visit to $z$, relative to paying the long-run rate $\rho^*$ at
each step.
Since $\hitz=0$ under $\Prob_z$, the empty sum gives $h^*(z)=0$. Moreover, by the triangle inequality and~\eqref{eq:G},
\begin{equation*}
  |h^*(x)|
  \le
  \E_x\Bigl[
    \sum_{t=0}^{\hitz-1}|c(X_t)-\rho^*|
  \Bigr]
  =
  G(x)
  <\infty.
\end{equation*}
Hence the canonical bias is well defined and finite at every state.

This first-passage representation differs from the classical interpretation of the bias for finite chains in~\cite[Sec.~8.2.1]{Puterman94}.
There, for an aperiodic chain, the bias of a state $x$ is the limit, as $N$ grows, of the difference between the expected costs accumulated in the first $N$ steps by the chain started at $x$ and by the chain started in steady state.
The canonical bias compares with the chain started at the reference state $z$ instead.
For a finite irreducible chain, $h^*$ is the bias of~\cite{Puterman94} shifted by a constant such that it is zero at $z$.

\begin{lemma}[Poisson equation]
\label{lem:poisson}
Under Assumption~\ref{ass:model},
\begin{equation}
\label{eq:poisson}
  \rho^*+h^*(x)=c(x)+(Ph^*)(x)\qquad\text{for every }x\in\States.
\end{equation}
\end{lemma}

\begin{example}[Canonical bias for the single-server queue]
\label{ex:queue-canonical}
Take the reference state $z=0$.
Since $c(X_t)=0$ for every $t<\sigma_0$, we obtain,
for $x\ge1$,
\[
  h^*(x)
  =
  \E_x\Bigl[
    \sum_{t=0}^{\sigma_0-1}
    \bigl(c(X_t)-\rho^*\bigr)
  \Bigr]
  =
  \E_x[-\rho^*\sigma_0]
  =
  -\rho^*\E_x[\sigma_0].
\]
For $x=0$, we have $\sigma_0=0$,
and $h^*(0)=0$.

Let $m(x):=\E_x[\tau_0]$ be the expected time to the
next visit to $0$ at a time $t\ge1$.
For $x\ge1$, we have $\tau_0=\sigma_0$, and conditioning
on the first transition gives
\begin{align*}
  m(1)&=1+\alpha m(2),\\
  m(x)&=1+\alpha m(x+1)+\beta m(x-1),
  \qquad x\ge2.
\end{align*}
The expected hitting time is the minimal nonnegative
solution of these equations
\cite[Sec.~1.3]{Norris97}, namely
$m(x)=x/(\beta-\alpha)$ for $x\ge1$.
For $x=0$, the first identity in~\eqref{eq:kac} gives
$m(0)=1/\pi(0)=\beta/(\beta-\alpha)$.
Consequently,
$
  h^*(x)=-\rho^*m(x)=-\frac{x}{\beta}, $ 
  for $ x\ge1,
$
and $h^*(0)=0$.
Thus, the bias used in Example~\ref{ex:cert}
is the canonical bias when $\theta=0$.

Fig.~\ref{fig:bias} shows the expected cost
$\E_x[\sum_{t=0}^{N-1}c(X_t)]$ accumulated in the first
$N$ steps for three initial states $x$.
For the chain started in steady state, this cost is
$N\rho^*$.
Each curve approaches a line with slope $\rho^*$,
shown dashed in the figure.
The dashed asymptote for initial state $x$ has an
intercept that differs from that for initial state $0$
by $h^*(x)$. These are the intercepts of the asymptotes;
the actual cost curves all start at zero because no
cost has accumulated at $N=0$.
Thus, in the long run, starting from $x$ changes the
expected accumulated cost by $h^*(x)$ compared with
starting from $0$. 

\begin{figure}[t]
\centering
\includegraphics[width=0.9\columnwidth]{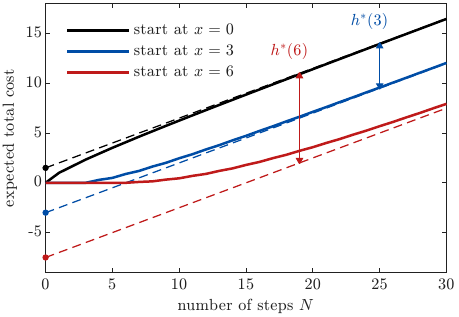}
\caption{Expected total cost of the queue for
$\alpha=1/3$ and initial states $0$, $3$ and $6$.
The dashed lines are the asymptotes, with slope
$\rho^*=1/2$ and intercepts $1.5$, $-3$ and $-7.5$.
The arrows mark the differences from the asymptote
of the reference state $0$, which are
$h^*(3)=-4.5$ and $h^*(6)=-9$.}
\label{fig:bias}
\end{figure}
\end{example}
Omitting the growth conditions (C2) and (D2) causes
unsoundness, as the following example demonstrates.

\begin{example}
\label{ex:false}
For the queue of Example~\ref{ex:queue}, fix
$\rho>\rho^*$ and let
\[
  h(x)=
  -\frac{\rho x}{\beta-\alpha}
  +b\left(\frac{\beta}{\alpha}\right)^x,
  \qquad
  b=\frac{\beta(\rho-\rho^*)}{(\beta-\alpha)^2}.
\]
For $x\ge1$, a direct calculation gives
$(Ph)(x)-h(x)=\rho$.
At $x=0$, we obtain
\[
  (Ph)(0)-h(0)
  =
  -\frac{\alpha\rho}{\beta-\alpha}
  +b(\beta-\alpha)
  =
  \rho-1.
\]
Since $c(0)=1$ and $c(x)=0$ for $x\ge1$,
the Poisson equation
\[
  \rho+h(x)=c(x)+(Ph)(x)
\]
holds at every state.
Thus, condition~\ref{itm:c1} holds with equality,
even though $\rho>\rho^*$ is not a valid lower bound.

The integrability clause of~\ref{itm:c2} also holds.
Indeed, $X_t\le x+t$ almost surely, so
\[
  |h(X_t)|
  \le
  \frac{|\rho|(x+t)}{\beta-\alpha}
  +|b|\left(\frac{\beta}{\alpha}\right)^{x+t}.
\]
The right-hand side is a finite deterministic bound
for every fixed $x$ and $t$.
Hence $\E_x[|h(X_t)|]<\infty$.

Taking expectations in the Poisson equation and summing
over $t=0,\dots,N-1$ gives
\[
  \frac{1}{N}\E_x[h(X_N)]
  =
  \frac{h(x)}{N}+\rho
  -\frac{1}{N}\E_x\Bigl[
    \sum_{t=0}^{N-1}c(X_t)
  \Bigr].
\]
By Lemma~\ref{lem:cesaro}, the expected average cost
converges to $\rho^*$.
Therefore,
\[
  \lim_{N\to\infty}
  \frac{1}{N}\E_x[h(X_N)]
  =
  \rho-\rho^*>0.
\]
Thus, the limit clause of~\ref{itm:c2} fails.
This example shows why the Poisson inequality and
finite-time integrability alone are insufficient.
\end{example}

The exponential component $\psi(x)=(\beta/\alpha)^x$
in the above example satisfies $(P\psi)(x)=\psi(x)$ at every state $x\ge1$. Thus, the above example shows that local one-step conditions can coexist with positive asymptotic growth.
The next section isolates this problem exactly with an asymptotic condition on an auxiliary bounding function. 

\section{The Lyapunov Rule}
\label{sec:fl}
The verification theorem of Section~\ref{sec:rule} is sound, but its growth
condition on $\E_x[h(X_N)]$ is not local and can be difficult to verify
directly. In this section, we control the growth of the bias by an
auxiliary Lyapunov function for the killed chain. This replaces the
growth condition by a condition on the Lyapunov function and prepares the
final reduction to locally checkable conditions.

\subsection{The Rule and Its Soundness}

\begin{definition}[Lyapunov certificate]
\label{def:fl}
A \emph{Lyapunov lower-bound certificate} consists of $\rho\in\R$, a function $h:\States\to\R$, a reference state $z\in\States$ and a function $w:\States\to[0,\infty)$ such that
\begin{condF}
\item \label{itm:f1} $\rho+h(x)\le c(x)+(Ph)(x)$ for every $x\in\States$,
\item \label{itm:f2} $(\taboo w)(x)\le w(x)$ for every $x\in\States$,
\item \label{itm:f3} $|h(x)|\le w(x)$ for every $x\in\States$,
\item \label{itm:f4} $(\taboo^{\,n}w)(z)\to0$ as $n\to\infty$.
\end{condF}
A \emph{Lyapunov upper-bound certificate} satisfies~\ref{itm:f1} with the reversed inequality and~\ref{itm:f2} to~\ref{itm:f4} unchanged.
\end{definition}

The reference state is now a part of the certificate, and the taboo kernel in~\ref{itm:f2} and~\ref{itm:f4} is defined with respect to that state.
We call $w$ the \emph{Lyapunov function},~\ref{itm:f2} the superharmonicity condition, \ref{itm:f3} the domination condition, and~\ref{itm:f4} the decay condition at $z$.
By $w_\infty=\lim_{n\to\infty}\taboo^{\,n}w$, this decay condition~\ref{itm:f4} states that the harmonic part of the Lyapunov function is zero at $z$.
Conditions~\ref{itm:f2} and~\ref{itm:f3} make $Ph$ well defined, because
\begin{align*}
  (P|h|)(x)
  \le
  \sum_y p(y\mid x)w(y)
  &=
  (\taboo w)(x)+p(z\mid x)w(z)
  \\ &\le
  w(x)+w(z)
  <\infty.
\end{align*}

\begin{theorem}[Soundness of the Lyapunov rule]
\label{thm:reduction}
Every Lyapunov lower-bound certificate $(\rho,h,z,w)$ gives a lower-bound certificate $(\rho,h)$, and every Lyapunov upper-bound certificate gives an upper-bound certificate.
Hence Theorem~\ref{thm:expected} applies.
\end{theorem}

The proof is given in Appendix~\ref{app:reduction}.
It rests on the last-exit bound
\begin{equation}
\label{eq:lastexit}
  \E_x[w(X_t)]
  \le
  (\taboo^{\,t}w)(x)
  +
  \sum_{n=0}^{t-1}(\taboo^{\,n}w)(z),
\end{equation}
which holds for every $x\in\States$ and $t\ge1$ and follows by splitting the trajectories according to their last visit to $z$ up to time $t$.

Eq.~\eqref{eq:lastexit} also explains why the decay condition is
required only at the reference state $z$.
The contribution associated with the initial state is bounded by $w(x)$ and therefore tends to zero after division by $t$.
In contrast, every visit to $z$ starts a new path to the next visit to $z$, and these paths produce the sum
$  \sum_{n=0}^{t-1}(\taboo^{\,n}w)(z)$.
The decay condition at $z$ ensures that the average
$\frac1t\sum_{n=0}^{t-1}(\taboo^{\,n}w)(z)$
tends to zero.

The use of the taboo kernel is also essential for allowing unbounded biases.
Under Assumption~\ref{ass:model}, if~\ref{itm:f2} were replaced by $Pw\le w$, then $w$ would be constant, because every nonnegative superharmonic function of an irreducible recurrent chain is constant~\cite[Prop.~6.3]{KemenySnellKnapp76}.
Together with $|h|\le w$, this would force $h$ to be bounded. 
Killing the chain at $z$ avoids this restriction and permits unbounded Lyapunov functions.

\subsection{Completeness}
\label{sec:flcomplete}

To prove completeness, we construct the Lyapunov function as the expected
accumulated absolute recentered cost, plus one per step, up to the next visit to $z$.
The following lemma shows that any finite expected total of this form is automatically superharmonic and satisfies the decay
condition (F4).

\begin{lemma}[Expected totals]
\label{lem:totals}
Let $g:\States\to[0,\infty)$ and suppose that $w=\sum_{n=0}^{\infty}\taboo^{\,n}g$ is finite at every state.
Then $w=g+\taboo w$, and $w$ satisfies~\ref{itm:f2} and~\ref{itm:f4}.
\end{lemma}

\begin{proof}
Tonelli's theorem gives $\taboo w=\sum_{n\ge1}\taboo^{\,n}g=w-g$.
Since $g\ge0$, this is~\ref{itm:f2}.
By the same theorem $(\taboo^{\,n}w)(z)=\sum_{k\ge n}(\taboo^{\,k}g)(z)$, which is the tail of a series with the finite sum $w(z)$ and tends to zero.
\end{proof}

\begin{definition}[Canonical Lyapunov function]
\label{def:wrho}
Under Assumption~\ref{ass:model} the \emph{canonical Lyapunov function} is
\begin{equation*}
  w^\dagger(x)=\E_x\Bigl[\sum_{t=0}^{\retz-1}\bigl(1+c'(X_t)\bigr)\Bigr],
\end{equation*}
where $c'$ is defined in~\eqref{eq:G}.
\end{definition}
Thus, $w^\dagger(x)$ accumulates the absolute recentered cost over the path to the next visit to $z$, and it additionally accounts for the duration of this path.
The additional duration enables a transformation required in the local rule of the next section.

\begin{theorem}[Completeness of the Lyapunov rule]
\label{thm:flcomplete}
Fix the reference state $z\in\States$ and let
Assumption~\ref{ass:model} hold. 
Then $w^\dagger(x)$ is finite at every state and satisfies
\begin{equation}
\label{eq:wident}
w^\dagger(x)=1+c'(x)+(\taboo w^\dagger)(x),
\qquad x\in\States.
\end{equation}
For every $\rho\le\rho^*$ the tuple $(\rho,h^*,z,w^\dagger)$ is a Lyapunov lower-bound certificate.
For every $\rho\ge\rho^*$ it is a Lyapunov upper-bound certificate.
\end{theorem}

\begin{proof}
For $x\ne z$ we have $\retz=\hitz$ under $\Prob_x$, and hence $w^\dagger(x)=\E_x[\hitz]+G(x)$, which is finite by Lemma~\ref{lem:hitting}.
At $z$, \eqref{eq:kac} gives $w^\dagger(z)=(1+\pi(c'))/\pi(z)$.
By~\eqref{eq:expected-total}, $w^\dagger=\sum_n\taboo^{\,n}(1+c')$.
Lemma~\ref{lem:totals} gives~\eqref{eq:wident}, \ref{itm:f2} and~\ref{itm:f4}.
For~\ref{itm:f3}, $|h^*(x)|\le G(x)\le w^\dagger(x)$ for $x\ne z$ and $h^*(z)=0$, and hence $|h^*|\le w^\dagger$.
Condition~\ref{itm:f1} follows from Lemma~\ref{lem:poisson}, since $\rho\le\rho^*$ gives $\rho+h^*\le\rho^*+h^*=c+Ph^*$, and its reversed form follows in the same way for $\rho\ge\rho^*$.
\end{proof}

By Theorem~\ref{thm:reduction}, the canonical bias therefore satisfies the growth conditions~\ref{itm:c2} and~\ref{itm:d2}, and $(\rho,h^*)$ is a lower-bound certificate for every $\rho\le\rho^*$ and an upper-bound certificate for every $\rho\ge\rho^*$.

\begin{example}[Lyapunov function for the single-server queue]
\label{ex:witness}
For the queue with $z=0$, the function
$m(x)=\E_x[\tau_0]$ of Example~\ref{ex:queue-canonical}
is the expected accumulated cost for the constant
cost function $g=1$. Lemma~\ref{lem:totals} gives $\taboo m=m-1$ and the decay condition at $0$.
Domination holds because $|h^*(x)|=x/\beta\le x/(\beta-\alpha)=m(x)$ for $x\ge1$ and $h^*(0)=0$.
Hence $(\rho,h^*,0,m)$ is a Lyapunov lower-bound certificate for every $\rho\le\rho^*$.
The function $e(x)=(\beta/\alpha)^x-1$ for $x\ge1$ with $e(0)=\beta-\alpha$ is harmonic for $\taboo$.
The function $m+e$ satisfies~\ref{itm:f2} and~\ref{itm:f3} with the same decrement as $m$ and violates~\ref{itm:f4}, since $(\taboo^{\,n}(m+e))(0)\to\beta-\alpha$. The decay condition (F4) is necessary for soundness.
Take $\rho>\rho^*$ and $h$ and $b$ from Example~\ref{ex:false}.
The function $w=b\,(e+1)+\rho\,m$ is nonnegative and satisfies
\begin{equation*}
  \taboo w=b\,e+b\,\taboo\ind_{\States}+\rho\,(m-1)\le w.
\end{equation*}
It dominates the bias, since $|h(x)|\le b(\beta/\alpha)^x+\rho x/(\beta-\alpha)=w(x)$ for $x\ge1$ and $|h(0)|=b\le w(0)$.
Hence $(\rho,h,0,w)$ satisfies~\ref{itm:f1}, \ref{itm:f2} and~\ref{itm:f3} for a bound that is false.
Only~\ref{itm:f4} fails, since $(\taboo^{\,n}w)(0)\to b\,(\beta-\alpha)>0$.
The harmonic part of this $w$ is $w_\infty=b\,e$, since $\taboo^{\,n}\ind_{\States}\to0$ and $\taboo^{\,n}m\to0$, and it comes from the exponential component $b(\beta/\alpha)^x$ of the false bias in Example~\ref{ex:false}.
The growth condition rules out this component in Section~\ref{sec:rule}, and the decay condition rules it out here.
\end{example}

The functions $m$ and $m+e$ of Example~\ref{ex:witness} have the same decrements.
Thus, the decrement \(w-\taboo w\) alone cannot distinguish a Lyapunov function from one with an added harmonic component.
Moreover, the false bound of Example~\ref{ex:false}, with the function $w$ of Example~\ref{ex:witness}, shows that omitting the decay condition makes the rule unsound.

The next section alters the chain using the Lyapunov function $w$ and turns the remaining decay condition~\ref{itm:f4} into a reachability property.

\section{The Local Rule}
\label{sec:local}
The Lyapunov rule reduces the growth condition to a decay condition on the
Lyapunov function, but this condition still involves a limit and is therefore not
local. In this section, we eliminate this final remaining nonlocal condition.
We transform the killed chain by the Lyapunov function, interpret the decay condition as a reachability property, and certify that property by local inequalities.
This leads to a fully local rule, for which we establish both soundness and
completeness.

\subsection{The Transformed Chain}
\label{sec:twist}

Throughout this subsection $w:\States\to(0,\infty)$ satisfies~\ref{itm:f2} for the reference state $z$.

\begin{definition}[Transformed chain]
\label{def:twist}
Let $\partial\notin\States$ be an additional state.
The \emph{$w$-transform} of the killed chain, which we call the \emph{transformed chain}, is the Markov chain with the state space $\States\cup\{\partial\}$ and the kernel
\begin{align*}
  K(x,y)&=\frac{p(y\mid x)\,w(y)}{w(x)},\qquad y\in\States\setminus\{z\},\\
  K(x,\partial)&=1-\frac{(\taboo w)(x)}{w(x)},
\end{align*}
for $x\in\States$, with $K(x,z)=0$ and $K(\partial,\partial)=1$.
\end{definition}

By~\ref{itm:f2} and the positivity of $w$, we have
$0\le(\taboo w)(x)/w(x)\le1$. Moreover,
$
  \sum_{y\ne z}K(x,y)
  =
  \frac{(\taboo w)(x)}{w(x)}$. Hence the remaining probability mass is exactly
$K(x,\partial)=1-(\taboo w)(x)/w(x)$, $K$ is a stochastic kernel
and $\partial$ is an absorbing state. The quantity
$K(x,\partial)=(w(x)-(\taboo w)(x))/w(x)$ is the probability that the transformed chain is sent from $x$ to $\partial$, and we call it the
\emph{killing probability} at $x$. It is the one-step decrement
of the Lyapunov function relative to its current value. Thus, when $w$ is harmonic at $x$, no
probability mass is sent to $\partial$, and when $w$ decreases at $x$, the
killing probability is positive.

\begin{remark}[Doob transform]
\label{rem:doob}
The construction in Definition~\ref{def:twist} is the $h$-transform of Doob~\cite{Doob84,KemenySnellKnapp76}, applied to the taboo kernel with the Lyapunov function in the role of $h$, and with the missing mass sent
to $\partial$. Its effect along a path is simple. For
$y_0,\dots,y_n$ with $y_1,\dots,y_n\ne z$,
\begin{equation*}
  \prod_{i=0}^{n-1}K(y_i,y_{i+1})
  =
  \frac{w(y_n)}{w(y_0)}
  \prod_{i=0}^{n-1}p(y_{i+1}\mid y_i).
\end{equation*}
Indeed, the intermediate factors of $w$ cancel. Hence the kernel $K$ reweights each surviving path by the ratio of the Lyapunov function at
its endpoint to its value at the starting state.
\end{remark}

\begin{proposition}[Survival identity]
\label{prop:twist}
Let $w>0$ satisfy~\ref{itm:f2}.
Then
\begin{romanlist}
\item for every $x\in\States$ and $n\in\N$, the transformed chain started at $x$ is not absorbed by time $n$ with probability $(\taboo^{\,n}w)(x)/w(x)$,
\item the transformed chain started at $x$ is never absorbed with probability $w_\infty(x)/w(x)$,
\item condition~\ref{itm:f4} holds if and only if the transformed chain started at $z$ reaches $\partial$ almost surely.
\end{romanlist}
\end{proposition}

\begin{proof}
We prove (i) by induction.
For $n=0$ both sides are one.
Conditioning on the first step and using the statement for $n$ gives
\begin{equation*}
  \sum_{y\ne z}\frac{p(y\mid x)w(y)}{w(x)}\frac{(\taboo^{\,n}w)(y)}{w(y)}=\frac{(\taboo^{\,n+1}w)(x)}{w(x)},
\end{equation*}
which is the statement for $n+1$.
The events in (i) decrease with $n$, and hence continuity from above and the definition of $w_\infty$ give (ii).
Statement (iii) is (ii) at $x=z$, since~\ref{itm:f4} states $w_\infty(z)=0$.
\end{proof}

After $n$ steps, $(\taboo^{\,n}w)(x)/w(x)$ is the probability that the
transformed chain has survived, while $w_\infty(x)/w(x)$ is the probability
that it survives forever.
Thus, the harmonic part of the Lyapunov function corresponds precisely to the probability mass that is never absorbed.
In this way, the transformed chain turns the decay condition into an absorption property.

\begin{example}[$w$-transformed chain of the single-server queue]
\label{ex:twisted}
We transform the queue by the Lyapunov function $m$ of Example~\ref{ex:witness}.
Since $\taboo m=m-1$, the killing probability is $K(x,\partial)=1/m(x)$, which is $(\beta-\alpha)/x$ at $x\ge1$ and $\rho^*$ at $0$.
The other entries are $K(x,x+1)=\alpha(x+1)/x$ and $K(x,x-1)=\beta(x-1)/x$ for $x\ge1$, and $K(0,1)=\alpha/\beta$.
By~\eqref{eq:taboo-iter} and the Markov property, $(\taboo^{\,n}m)(0)=\E_0[\max\{\tau_0-n,0\}]$.
Proposition~\ref{prop:twist} gives the survival probability
\begin{equation*}
  \frac{(\taboo^{\,n}m)(0)}{m(0)}=\frac{\E_0[\max\{\tau_0-n,0\}]}{\E_0[\tau_0]},
\end{equation*}
which tends to zero by the dominated convergence theorem.
The transformed chain started at $0$ reaches $\partial$ almost surely.
Fig.~\ref{fig:twisted} shows the transformed chain for $\alpha=1/3$.

\begin{figure}[t]
\centering
\begin{tikzpicture}
  \queuenodes
  \foreach \a/\b/\l in {n0/n1/{1/2},n1/n2/{2/3},n2/n3/{1/2},n3/n4/{4/9},n4/nd/{5/12}} \draw[->] (\a) to[bend left=40] node[ev,above]{$\l$} (\b);
  \foreach \a/\b/\l/\q in {n2/n1/{1/3}/0.42,n3/n2/{4/9}/0.62,n4/n3/{1/2}/0.78,nd/n4/{8/15}/0.5} \draw[->] (\a) to[bend left=40] node[ev,below,pos=\q]{$\l$} (\b);
  \draw[del] (n1) to[bend left=40] (n0);
  \draw[del] (n0) to[loop left] (n0);
  \node[qabs] (d) at (2*\qdx,-2.8) {$\partial$};
  \draw[->] (d) to[loop right] node[ev,right]{$1$} (d);
  \foreach \a/\l in {n0/{1/2},n1/{1/3},n2/{1/6},n3/{1/9},n4/{1/12}} \draw[->] (\a) -- node[ev,fill=white,inner sep=1pt,pos=0.42]{$\l$} (d);
\end{tikzpicture}
\caption{Transformed chain of the queue for $\alpha=1/3$ and the Lyapunov function $m$. The transitions into the reference state $0$ are deleted and shown dashed. The reference state has a double border.}
\label{fig:twisted}
\end{figure}
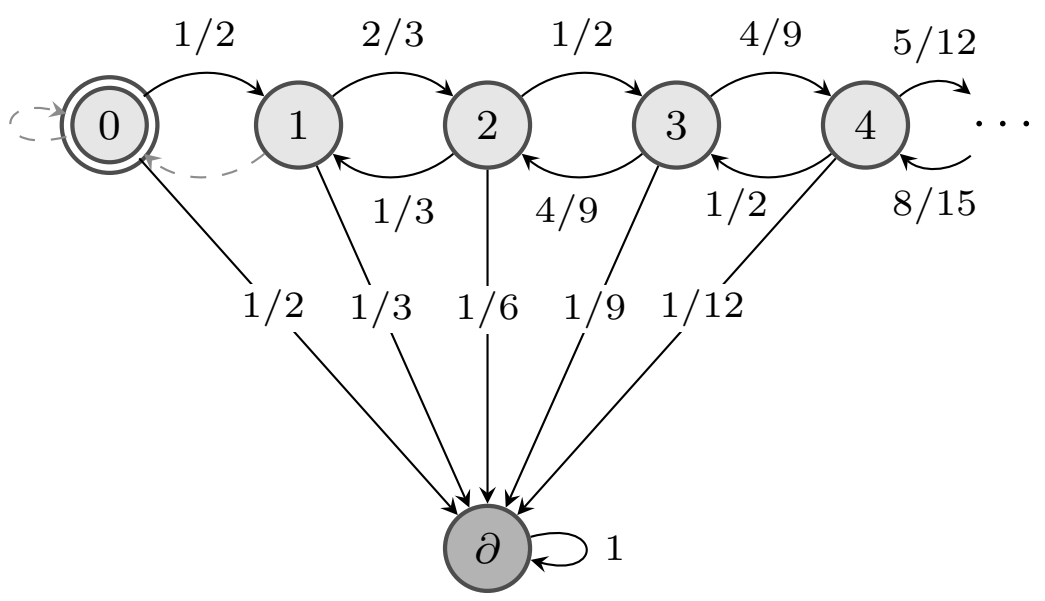
\end{example}

\subsection{Reachability Certificates}

Proposition~\ref{prop:twist} turns the decay condition into almost-sure reachability of $\partial$ in a countable Markov chain.
For this property a sound and complete rule is available.

\begin{definition}[Reachability certificate]
\label{def:reachcert}
Let $M$ be a Markov chain on a countable state space $\States_M$ with kernel $P_M$ and an absorbing state $\partial$, and let $s_0\in\States_M$.
A \emph{reachability certificate} for $M$ at $s_0$ consists of
\begin{condR}
\item \label{itm:r1} a set $\Inv\subseteq\States_M$ that contains $s_0$ and $\partial$ such that $s\in\Inv$ and $P_M(s,s')>0$ imply $s'\in\Inv$,
\item \label{itm:r2} a function $V:\Inv\to[0,\infty)$ with $V(s)=0$ if and only if $s=\partial$, and with $V(s)\ge\sum_{s'}P_M(s,s')V(s')$ for every $s\in\Inv$,
\item \label{itm:r3} a function $U:\Inv\to\N$ with $U(\partial)=0$ such that, for every $r\ge0$, $U$ is bounded on the set $\{s\in\Inv \mid V(s)\le r\}$ and there is $\varepsilon_r>0$ with $\sum_{s'\,:\,U(s')<U(s)}P_M(s,s')\ge\varepsilon_r$ for every $s\ne\partial$ in that set.
\end{condR}
\end{definition}

This is the martingale rule of~\cite{MS25}.
The set $\Inv$ confines the trajectories reachable from $s_0$.
The function $V$ is a nonnegative supermartingale that controls the region visited by the chain, and the integer-valued variant $U$ measures the progress toward $\partial$.
Condition~\ref{itm:r3} requires that, on every sublevel set of $V$, the variant is bounded and decreases with a probability that is positive uniformly over the states other than $\partial$.

\begin{theorem}[Martingale rule \cite{MS25}]
\label{thm:reach}
The chain $M$ reaches $\partial$ from $s_0$ almost surely if and only if there is a reachability certificate for $M$ at $s_0$.
If $M$ reaches $\partial$ from $s_0$ almost surely, then the certificate can be chosen such that the set $\{s\in\Inv \mid V(s)\le r\}$ is finite for every $r\ge0$.
\end{theorem}

Soundness combines the convergence theorem for nonnegative supermartingales with a bounded-variant argument on each sublevel set of $V$.
Completeness constructs $V$ from reachability probabilities as shown in~\cite{MSZ78} and the variant from the lengths of shortest paths to $\partial$.
The local rule depends on both directions of Theorem~\ref{thm:reach}.
To keep the paper self-contained, Appendix~\ref{app:reach} gives a proof.

\subsection{The Rule}
We apply Definition~\ref{def:reachcert} to the transformed chain started at $z$.
For this chain the variant decreases at $x$ either by absorption, which has the killing probability $K(x,\partial)$, or by a transition to a state $y\ne z$ with $U(y)<U(x)$.
Hence the probability of a decrease at $x$ is
\begin{equation*}
  K(x,\partial)+\sum_{\substack{y\ne z\\ U(y)<U(x)}}K(x,y)
  =
  1-\sum_{\substack{y\ne z\\ U(y)\ge U(x)}}\frac{p(y\mid x)\,w(y)}{w(x)}.
\end{equation*}
We write the bound and the rates of~\ref{itm:r3} as a nondecreasing and a nonincreasing function of the level of $V$, which turns both requirements into pointwise inequalities at each state.

\begin{definition}[Local certificate]
\label{def:local}
A \emph{local lower-bound certificate} consists of $\rho\in\R$, functions $h:\States\to\R$ and $w:\States\to(0,\infty)$, a reference state $z\in\States$, a set $\Inv\subseteq\States$ with $z\in\Inv$, functions $V:\Inv\to(0,\infty)$ and $U:\Inv\to\N$, a nondecreasing function $\nu:[0,\infty)\to\N$, and a nonincreasing function $\varepsilon:[0,\infty)\to(0,1]$ such that
\begin{condL}
\item \label{itm:l1} $\rho+h(x)\le c(x)+(Ph)(x)$ for every $x\in\States$,
\item \label{itm:l2} $(\taboo w)(x)\le w(x)$ for every $x\in\States$,
\item \label{itm:l3} $|h(x)|\le w(x)$ for every $x\in\States$,
\item \label{itm:l4} $x\in\Inv$, $y\ne z$ and $p(y\mid x)>0$ imply $y\in\Inv$,
\item \label{itm:l5} $w(x)V(x)\ge\sum_{y\ne z}p(y\mid x)\,w(y)V(y)$ for every $x\in\Inv$,
\item \label{itm:l6} $1\le U(x)\le\nu\bigl(V(x)\bigr)$ for every $x\in\Inv$,
\item \label{itm:l7} $\sum_{y\ne z,\ U(y)\ge U(x)}p(y\mid x)\,w(y)\le\bigl(1-\varepsilon(V(x))\bigr)\,w(x)$ for every $x\in\Inv$.
\end{condL}
A \emph{local upper-bound certificate} satisfies~\ref{itm:l1} with the reversed inequality and~\ref{itm:l2} to~\ref{itm:l7} unchanged.
\end{definition}

We call $V$ the \emph{supermartingale}, $U$ the \emph{variant}, $\nu$ the \emph{variant bound} and $\varepsilon$ the \emph{progress function}. The names of $V$ and $U$ follow~\cite{MS25}.
By~\ref{itm:l4} the sums in~\ref{itm:l5} and~\ref{itm:l7} run over states of $\Inv$.
Condition~\ref{itm:l5} states that the product $wV$ is superharmonic for $\taboo$ on $\Inv$.
Condition~\ref{itm:l6} bounds the variant on every sublevel set of $V$.
Condition~\ref{itm:l7} states that the transformed chain moves from $x$ to $\partial$ or to a state with a smaller value of $U$ with probability at least $\varepsilon(V(x))$.
Since $\varepsilon$ is nonincreasing, this probability is at least $\varepsilon(r)$ on the whole sublevel set of $V$ at the level $r$.

Each of the seven conditions is a local condition at a single state $x$.
It involves the values of $h$, $w$, $V$ and $U$ at $x$, the functions $\nu$ and $\varepsilon$ at $V(x)$, the probabilities $p(\cdot\mid x)$, and the values at the successors of $x$.
No condition contains a limit along the chain.
For the constant functions $V=1$ and $U=1$, condition~\ref{itm:l5} is~\ref{itm:l2}, condition~\ref{itm:l6} holds with $\nu=1$, and~\ref{itm:l7} reads $\taboo w\le(1-\varepsilon(1))\,w$ on $\Inv$.

\begin{theorem}[The local rule is sound and complete]
\label{thm:local}
Every local lower-bound certificate $(\rho,h,z,w,\Inv,V,U,\nu,\varepsilon)$ satisfies $J(x)\ge\rho$ for every $x\in\States$, and every local upper-bound certificate satisfies $\Jbar(x)\le\rho$ for every $x\in\States$.
Under Assumption~\ref{ass:model}, a local lower-bound certificate with the scalar $\rho$ exists if and only if $\rho\le\pi(c)$, and a local upper-bound certificate with the scalar $\rho$ exists if and only if $\rho\ge\pi(c)$.
\end{theorem}

\begin{proof}
We extend $V$ and $U$ by $V(\partial)=U(\partial)=0$ and show first that~\ref{itm:l4} to~\ref{itm:l7} state that $\Inv\cup\{\partial\}$, $V$ and $U$ form a reachability certificate for the transformed chain at $z$.
Since $w>0$, the successors of $x\in\Inv$ under $K$ are $\partial$ and the states $y\ne z$ with $p(y\mid x)>0$.
Hence~\ref{itm:l4} is~\ref{itm:r1}.
Division of~\ref{itm:l5} by $w(x)$ gives $V(x)\ge\sum_{y\ne z}K(x,y)V(y)$, which is~\ref{itm:r2} because $V(\partial)=0$ and $V>0$ on $\Inv$.
Let $r\ge0$ and $x\in\Inv$ with $V(x)\le r$.
By~\ref{itm:l6} and the monotonicity of $\nu$, $U(x)\le\nu(r)$.
By~\ref{itm:l7}, the transformed chain moves from $x$ to a state with a smaller value of $U$ with probability at least $\varepsilon(V(x))$, and $\varepsilon(V(x))\ge\varepsilon(r)$ by the monotonicity of $\varepsilon$.
Hence~\ref{itm:l6} and~\ref{itm:l7} give~\ref{itm:r3} with the bound $\nu(r)$ and $\varepsilon_r=\varepsilon(r)$.
Conversely, let $\Inv\cup\{\partial\}$, $V$ and $U$ form a reachability certificate for the transformed chain at $z$.
Then $U\ge1$ on $\Inv$, because~\ref{itm:r3} with $r=V(x)$ requires a decrease of $U$ with positive probability at every $x\in\Inv$.
Conditions~\ref{itm:l6} and~\ref{itm:l7} hold with $\nu(r)$ the largest value of $U$ on the sublevel set $\{x\in\Inv \mid V(x)\le r\}$, and with $\varepsilon(r)$ the minimum of one and the infimum over this set of the probability that $U$ decreases, both equal to one if the set is empty.
By~\ref{itm:r3}, $\varepsilon(r)\ge\min\{1,\varepsilon_r\}>0$.

\emph{Soundness.}
Theorem~\ref{thm:reach} shows that the transformed chain reaches $\partial$ from $z$ almost surely.
Proposition~\ref{prop:twist} gives~\ref{itm:f4}.
Conditions~\ref{itm:l1} to~\ref{itm:l3} are~\ref{itm:f1} to~\ref{itm:f3}.
Theorems~\ref{thm:reduction} and~\ref{thm:expected} give the bound.

\emph{Completeness.}
Let $\rho\le\rho^*$.
The tuple $(\rho,h^*,z,w^\dagger)$ of Theorem~\ref{thm:flcomplete} satisfies~\ref{itm:l1} to~\ref{itm:l3}, and $w^\dagger\ge1$.
By~\ref{itm:f4} and Proposition~\ref{prop:twist} its transformed chain reaches $\partial$ from $z$ almost surely.
Theorem~\ref{thm:reach} gives a reachability certificate at $z$, and the correspondence above gives~\ref{itm:l4} to~\ref{itm:l7} after $\partial$ is removed from the invariant.
Conversely, a certificate with the scalar $\rho$ gives $J\ge\rho$ by soundness, and $J=\pi(c)$ by Lemma~\ref{lem:cesaro}.
The proof for upper bounds is the same with the reversed inequality in~\ref{itm:l1}.
\end{proof}

Note that the completeness proof uses $w^\dagger\ge1$, which makes the transformed chain well defined.
Without the term that counts the duration, the Lyapunov function would vanish at a state $x$ when the cost equals $\rho^*$ on every state that the chain visits from $x$ before the time $\retz$.

Theorem~\ref{thm:local} solves Problem~\ref{prob:main}.
The limit that the growth condition and the decay condition carry is discharged once, inside the soundness proof of Theorem~\ref{thm:reach}, where it appears as the convergence of a nonnegative supermartingale.

\begin{example}[Local certificate for the single-server queue]
\label{ex:local}
For the queue let $\rho\le\rho^*$, $h=h^*$, $z=0$, $w=m$, $\Inv=\N$, $V(x)=x+a$ with $a=1/(\beta-\alpha)$, $U(x)=x+1$, $\nu(r)=\lfloor r\rfloor$ and $\varepsilon=\beta-\alpha$.
Then $m(x)=ax$ for $x\ge1$ and $m(0)=\beta a$.
Conditions~\ref{itm:l1} to~\ref{itm:l3} hold by Example~\ref{ex:witness} and~\ref{itm:l4} holds because $\Inv=\States$.
For~\ref{itm:l5} at $x\ge1$ we divide by $a$ and obtain
\begin{equation*}
  \alpha(x+1)(x+1+a)+\beta(x-1)(x-1+a)=x(x+a)-2x(\beta-\alpha),
\end{equation*}
which is at most $x(x+a)$.
At $x=0$ the sum has the single term $\alpha m(1)V(1)=\alpha a(1+a)$, and $m(0)V(0)-\alpha a(1+a)=a(1-\alpha)>0$.
Condition~\ref{itm:l6} holds because $a\ge1$ gives $U(x)=x+1\le\lfloor x+a\rfloor=\nu(V(x))$.
For~\ref{itm:l7}, the only successor $y\ne0$ of $x$ with $U(y)\ge U(x)$ is $x+1$, and $1-\varepsilon=2\alpha$.
Hence~\ref{itm:l7} reads $\alpha a(x+1)\le2\alpha ax$ at $x\ge1$ and $\alpha a\le2\alpha\beta a$ at $x=0$, which hold because $x\ge1$ and $\beta\ge\tfrac12$.
The tuple is a local lower-bound certificate for every $\rho\le\rho^*$. The same $h^*$, $w$, $\Inv$, $V$, $U$, $\nu$ and $\varepsilon$ give a local upper-bound certificate for every $\rho\ge\rho^*$, with the inequality in~\ref{itm:l1} reversed.
In the transformed chain of Example~\ref{ex:twisted}, the variant decreases at $x\ge2$ by a step to $x-1$ or to $\partial$, with probability $K(x,x-1)+K(x,\partial)=\beta-\alpha/x$, at $x=1$ by absorption, with probability $\beta-\alpha$, and at $0$ by absorption, with probability $\rho^*$.
Hence $\beta-\alpha$ is the largest constant progress function.
Fig.~\ref{fig:local} shows $U$ and these probabilities for $\alpha=1/3$.

\begin{figure}[t]
\centering
\begin{tikzpicture}
  \queuenodes
  \queueedges{$\alpha$}{$\beta$}
  \draw[->] (n1) to[bend left=40] node[ev,below]{$\beta$} (n0);
  \draw[->] (n0) to[loop left] node[ev,left]{$\beta$} (n0);
  \foreach \k/\v in {0/1,1/2,2/3,3/4,4/5} \node[ev] at (\k*\qdx,1.05) {$U=\v$};
  \foreach \k/\v in {0/{1/2},1/{1/3},2/{1/2},3/{5/9},4/{7/12}} \node[ev] at (\k*\qdx,-1.05) {$\v$};
  \node[qref,qsub] at (0,0) {$0$};
  \node[qn,qsub] at (\qdx,0) {$1$};
  \node[qn,qsub] at (2*\qdx,0) {$2$};
\end{tikzpicture}
\caption{Local certificate of the queue for $\alpha=1/3$. The variant $U$ is shown above each state and the probability that $U$ decreases in the transformed chain below it. The states of the sublevel set $\{V\le5\}$ of $V(x)=x+3$ are shaded. The reference state has a double border.}
\label{fig:local}
\end{figure}
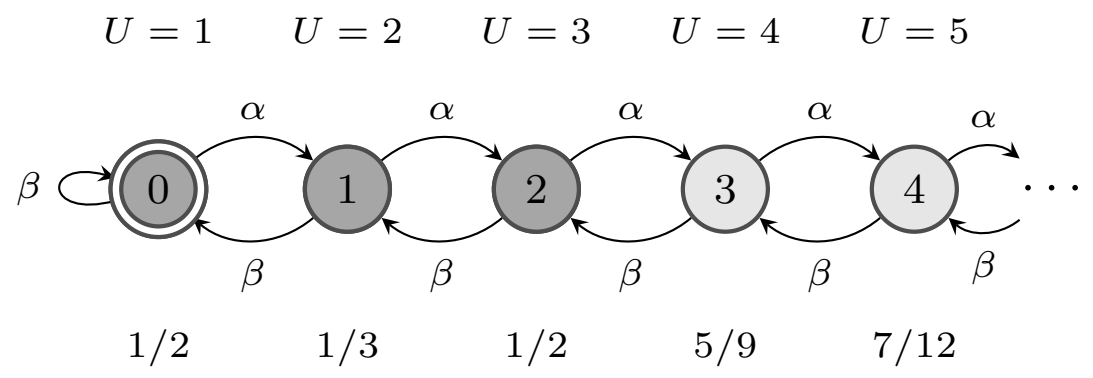
\end{example}

\begin{remark}[Reference state]
\label{rem:reference}
The reference state is a component of the certificate and is chosen by the designer.
Under Assumption~\ref{ass:model} every state can be chosen as the reference state, because the canonical objects $h^*$ and $w^\dagger$ are defined for every $z$.
Different reference states give different canonical biases and Lyapunov functions.
The extension of the Lyapunov condition from one reference state to a finite taboo set is studied in \cite[Sec.~5.2]{ABFGM93}.
\end{remark}

\begin{remark}[Finding certificates]
\label{rem:search}
For fixed choices or parametric families of the certificate functions,
the local conditions reduce the search for a certificate to a
feasibility problem over their parameters. In particular,
\ref{itm:l1} and~\ref{itm:l2} are linear in $(\rho,h,w)$, while
\ref{itm:l3} is equivalent to the two linear inequalities
$-w\le h\le w$. The remaining conditions depend on the chosen
parametrization of $V$, $U$, $\nu$ and $\varepsilon$ and need not be convex.
Any certificate found in a restricted family is sound, but failure
to find one does not invalidate the bound, since completeness is over
the unrestricted class of certificate functions.
For almost-sure reachability, sum-of-squares programs search such families for polynomial systems~\cite{KMS26}, and fixed polynomial or quadratic templates are known to lose completeness~\cite{KMMS26}.
\end{remark}

\section{Case Study: Networked Control over a Packet-Dropping Channel}
\label{sec:networked_control}

We apply the local rule to a control loop that receives the state over a lossy channel.
The plant is the stochastic linear system
\begin{equation*}
  \zeta_{t+1}=A\zeta_t+Bu_t+\omega_t,
\end{equation*}
where $\zeta_t\in\R^{d}$ is the state, $u_t\in\R^{d_u}$ is the input, and $(\omega_t)_{t\in\N}$ is an independent and identically distributed sequence with zero mean and covariance $W\succ0$.
At every step the state is delivered to the controller with probability $\beta\in(0,1)$, independently over time. Moreover, the packet-delivery process is independent of the process noise and the initial state.
We write $\alpha=1-\beta$ for the packet-loss probability.
A delivered packet reveals the state exactly.
Between deliveries the controller propagates its estimate through the model.
Under the certainty-equivalence information structure, the part of the quadratic cost that is caused by the estimation error is a function of the age of the last received state~\cite{YatesEtAl2021,ChampatiEtAl2019,KluegelEtAl2019}.
Let $I_t\in\N$ be the number of consecutive packet losses at time $t$.
The process $(I_t)_{t\in\N}$ is a Markov chain on $\States=\N$ with $p(0\mid i)=\beta$ and $p(i+1\mid i)=\alpha$ for every $i\in\N$, i.e., from the age $i$ the chain moves to $0$ with probability $\beta$ and to $i+1$ with probability $\alpha$.
It is irreducible and positive recurrent with the stationary distribution
\begin{equation}
\label{eq:age_stationary}
  \pi(i)=\beta\,\alpha^{\,i},\qquad i\in\N.
\end{equation}
After $i$ consecutive losses the covariance of the prediction error is
\begin{equation*}
  \Sigma_i=\sum_{j=0}^{i-1}A^jW(A^\top)^j,\qquad\Sigma_0=0,
\end{equation*}
and the cost is $c(i)=\tr(H\Sigma_i)$, where $H\succ0$ weights the estimation error.
In a linear quadratic Gaussian formulation, $H$ follows from the cost matrices and the associated Riccati equation, see, e.g.,~\cite[Sec.~4.2]{Bertsekas17}.
The matrix $A$ determines how fast an old state becomes costly, and $\beta$ determines the recurrence of the age.
The cost is unbounded when $A$ has an eigenvalue outside the open unit disc, and it grows exponentially in $i$ when $A$ has an eigenvalue outside the closed unit disc.

\subsection{Gain and Bias}

We assume
\begin{equation}
\label{eq:mean_square_condition}
  \alpha\,\sr(A)^2<1,
\end{equation}
where $\sr(A)$ is the spectral radius of $A$.
This condition is standard in remote estimation over packet-dropping channels~\cite{LiuEtAl2022}.
Under~\eqref{eq:mean_square_condition} the equation
\begin{equation}
\label{eq:discounted_lyapunov}
  X_\alpha=W+\alpha AX_\alpha A^\top
\end{equation}
has the unique solution $X_\alpha=\sum_{j=0}^{\infty}\alpha^{\,j}A^jW(A^\top)^j\succ0$.
By~\eqref{eq:age_stationary} and Tonelli's theorem,
\begin{align}
  \rho^*&=\sum_{i=0}^{\infty}\beta\,\alpha^{\,i}c(i)
  =\alpha\sum_{j=0}^{\infty}\alpha^{\,j}\tr\bigl(HA^jW(A^\top)^j\bigr)\notag\\
  &=\alpha\,\tr(HX_\alpha).
\label{eq:ncs_average_cost}
\end{align}
Hence $\pi(|c|)<\infty$ and Assumption~\ref{ass:model} holds.
Since $W\succ0$ and $H\succ0$, condition~\eqref{eq:mean_square_condition} is also necessary for $\rho^*<\infty$.

We take the reference state $z=0$.
From the state $i$ the chain visits $i,i+1,i+2,\dots$ until the first delivery, and it is at $i+t$ at time $t$ with probability $\alpha^{\,t}$ before that delivery.
Let
\begin{equation*}
  F(i)=\sum_{t=0}^{\infty}\alpha^{\,t}c(i+t),\qquad i\in\N.
\end{equation*}
The identity $\Sigma_{i+t}=\Sigma_i+A^i\Sigma_t(A^\top)^i$ and~\eqref{eq:discounted_lyapunov} give the closed form
\begin{equation}
\label{eq:ncs_F_closed}
  F(i)=\frac1\beta\Bigl[c(i)+\alpha\,\tr\bigl(HA^iX_\alpha(A^\top)^i\bigr)\Bigr].
\end{equation}
Since $\E_i[\hitz]=1/\beta$ for $i\ge1$ and $\rho^*=\beta F(0)$, the canonical bias of Definition~\ref{def:canonical} is
\begin{equation}
\label{eq:ncs_bias_general}
  h^*(i)=F(i)-\frac{\rho^*}{\beta}=F(i)-F(0).
\end{equation}
The taboo kernel at $0$ is $(\taboo f)(i)=\alpha f(i+1)$, and $h^*(0)=0$ gives $(Ph^*)(i)=\alpha h^*(i+1)$.
The Poisson equation~\eqref{eq:poisson} therefore reads
\begin{equation}
\label{eq:ncs_poisson}
  \rho^*+h^*(i)=c(i)+\alpha\,h^*(i+1),\qquad i\in\N,
\end{equation}
which also follows directly from $F(i)=c(i)+\alpha F(i+1)$.
The cost is nondecreasing in $i$. Hence, $h^*\ge0$, and the bias is unbounded whenever the cost is.

\subsection{Local Certificate}

We construct a Lyapunov function from a second Lyapunov equation.
We choose
\begin{equation}
\label{eq:eta_choice}
  \max\{1,\sr(A)^2\}<\eta<\frac1\alpha,
\end{equation}
which is possible by~\eqref{eq:mean_square_condition}, and let $Y\succ0$ solve
\begin{equation}
\label{eq:S_lyapunov}
  Y-\eta^{-1}A^\top YA=I.
\end{equation}
Let $\kappa_H$ be the largest eigenvalue of $Y^{-1/2}HY^{-1/2}$ and define
\begin{equation}
\label{eq:ncs_w_general}
  \overline C=\frac{\kappa_H\tr(YW)}{\eta-1},\qquad
  w(i)=\frac{\overline C}{1-\alpha\eta}\,\eta^{\,i}.
\end{equation}

\begin{proposition}[Certificate for the networked control system]
\label{prop:ncs}
Let~\eqref{eq:mean_square_condition} hold, let $\eta$, $Y$ and $w$ be as in~\eqref{eq:eta_choice} to~\eqref{eq:ncs_w_general}, and let $\Inv=\N$, $V=1$, $U=1$, $\nu=1$ and $\varepsilon=1-\alpha\eta$.
Then $(\rho,h^*,0,w,\Inv,V,U,\nu,\varepsilon)$ is a local lower-bound certificate for every $\rho\le\rho^*$ and a local upper-bound certificate for every $\rho\ge\rho^*$.
The killing probability of the transformed chain is $1-\alpha\eta$ at every state.
\end{proposition}

\begin{proof}
Equation~\eqref{eq:ncs_poisson} gives~\ref{itm:l1} for $\rho\le\rho^*$ and its reversed form for $\rho\ge\rho^*$.
The Lyapunov function satisfies $(\taboo w)(i)=\alpha w(i+1)=\alpha\eta\,w(i)$, and $\alpha\eta<1$ gives~\ref{itm:l2}.
For~\ref{itm:l3}, we have $H\preceq\kappa_HY$, and~\eqref{eq:S_lyapunov} gives $A^\top YA=\eta(Y-I)\preceq\eta Y$.
Hence $\tr\bigl(HA^jW(A^\top)^j\bigr)\le\kappa_H\eta^{\,j}\tr(YW)$ and
\begin{equation}
\label{eq:ncs_growth_bounds}
  c(i)\le\overline C\,(\eta^{\,i}-1),\qquad
  F(i)\le\frac{\overline C}{1-\alpha\eta}\,\eta^{\,i}.
\end{equation}
Since $0\le h^*\le F$, this gives $|h^*|\le w$.
Condition~\ref{itm:l4} holds because $\Inv=\N$.
With $V=1$, condition~\ref{itm:l5} is~\ref{itm:l2}, and~\ref{itm:l6} holds with $U=\nu=1$.
For~\ref{itm:l7}, the only successor $y\ne0$ of $i$ is $i+1$, and $U(i+1)=U(i)$.
Hence the left side of~\ref{itm:l7} is $\alpha w(i+1)=\alpha\eta\,w(i)$, the killing probability is
\begin{equation}
\label{eq:ncs_relative_decrement}
  \frac{w(i)-(\taboo w)(i)}{w(i)}=1-\alpha\eta>0
\end{equation}
at every $i\in\N$, and the constant function $\varepsilon=1-\alpha\eta$ satisfies~\ref{itm:l7}.
\end{proof}

The certificate holds on the whole countable state space and requires no truncation.
The sublevel sets of $V$ are not finite here.
The uniform bound~\eqref{eq:ncs_relative_decrement} makes~\ref{itm:l7} hold without them.
With the constant supermartingale and the constant variant, \ref{itm:l5} to~\ref{itm:l7} reduce to the geometric drift inequality $\taboo w\le\alpha\eta\,w$ for the killed chain.
The Lyapunov function grows at the rate $\eta$, which is
chosen between $\max\{1,\sr(A)^2\}$ and $1/\alpha$.
Such a rate exists if and only if~\eqref{eq:mean_square_condition} holds.
The same functions $h^*$, $w$, $V$ and $U$ certify every valid bound in both directions, and only the direction of~\ref{itm:l1} changes.

\subsection{Bounds from Inexact Certificates}

A certificate does not need the canonical bias, and a bound does not need the gain.
The next result certifies a lower and an upper bound on $\rho^*$ without the solution of~\eqref{eq:discounted_lyapunov}.
For $n\ge1$ let
\begin{equation}
\label{eq:inexact_bounds}
  \underline\rho_n=\beta\sum_{i=0}^{n-1}\alpha^{\,i}c(i)+\alpha^{\,n}c(n),\qquad
  \overline\rho=\frac{\alpha\,\kappa_H\tr(YW)}{1-\alpha\eta}.
\end{equation}

\begin{proposition}[Two-sided bound]
\label{prop:inexact}
Let~\eqref{eq:mean_square_condition} hold and let $\eta$, $Y$ and $w$ be as in Proposition~\ref{prop:ncs}.
Then, for every $n\ge1$, there are local certificates for the lower bound $\underline\rho_n$ and for the upper bound $\overline\rho$.
Hence $\underline\rho_n\le\rho^*\le\overline\rho$, and $\underline\rho_n$ increases to $\rho^*$ as $n\to\infty$.
\end{proposition}

\begin{proof}
\emph{Upper bound.}
Let $h(i)=\overline C\,(\eta^{\,i}-1)/(1-\alpha\eta)$.
Then $0\le h\le w$ and $h(0)=0$, and~\eqref{eq:ncs_growth_bounds} gives
\begin{equation*}
  c(i)+\alpha\,h(i+1)-h(i)\le\overline C\,(\eta^{\,i}-1)+\frac{\beta\,\overline C}{1-\alpha\eta}-\overline C\,\eta^{\,i}=\overline\rho.
\end{equation*}
This is~\ref{itm:l1} with the reversed inequality.
Conditions~\ref{itm:l2} to~\ref{itm:l7} hold as in Proposition~\ref{prop:ncs}.

\emph{Lower bound.}
Let $c_n(i)=c(\min\{i,n\})$.
The cost is nondecreasing, hence $c_n\le c$, and $\underline\rho_n=\pi(c_n)$.
Let $F_n(i)=\sum_{t\ge0}\alpha^{\,t}c_n(i+t)$ and $h_n=F_n-F_n(0)$.
As in~\eqref{eq:ncs_poisson}, $\underline\rho_n+h_n(i)=c_n(i)+\alpha\,h_n(i+1)\le c(i)+\alpha\,h_n(i+1)$, which is~\ref{itm:l1}.
The bias satisfies $0\le h_n\le c(n)/\beta$.
The constant Lyapunov function $w_n=c(n)/\beta$ is positive, satisfies $\taboo w_n=\alpha\,w_n$, and gives the killing probability $\beta$ at every state.
With $\Inv=\N$, $V=U=\nu=1$ and $\varepsilon=\beta$ this gives~\ref{itm:l2} to~\ref{itm:l7}.
Theorem~\ref{thm:local} gives both bounds, and monotone convergence gives $\underline\rho_n\to\pi(c)$.
\end{proof}

The lower bound in~\eqref{eq:inexact_bounds} is a finite sum and its certificate is bounded.
The upper bound uses the solution of~\eqref{eq:S_lyapunov} only, and its certificate grows at the rate $\eta$. When the cost is unbounded above, an upper-bound certificate requires an unbounded dominating function, whereas the truncated lower-bound certificates above have bounded dominating functions. A lower bound for a cost that is bounded below follows from a truncation of the cost, as in the proof above, while an upper bound must control the cost at every state.

\subsection{One-Dimensional System}

Let $d=1$ and $A=a$ with $|a|>1$, and write $\gamma=HW>0$.
The cost is
\begin{equation*}
  c(i)=\gamma\sum_{j=0}^{i-1}a^{2j}=\gamma\,\frac{a^{2i}-1}{a^2-1}.
\end{equation*}
If $\alpha a^2<1$, then $X_\alpha=W/(1-\alpha a^2)$, and~\eqref{eq:ncs_average_cost} and~\eqref{eq:ncs_bias_general} give
\begin{equation*}
  \rho^*=\frac{\gamma\,\alpha}{1-\alpha a^2},\qquad
  h^*(i)=\frac{\gamma\,(a^{2i}-1)}{(a^2-1)(1-\alpha a^2)}.
\end{equation*}
In the one-dimensional case the Lyapunov function can grow at the rate $a^2$ of the cost itself.
The function
\begin{equation*}
  w(i)=\frac{\gamma\,a^{2i}}{(a^2-1)(1-\alpha a^2)}
\end{equation*}
satisfies $|h^*|\le w$ and $(\taboo w)(i)=\alpha a^2w(i)$.
With $\Inv=\N$ and $V=U=1$ the killing probability is $1-\alpha a^2$ at every state.
A local certificate for a finite upper bound exists exactly when
\begin{equation}
\label{eq:critical}
  \beta>1-\frac{1}{a^2},
\end{equation}
which is the known critical arrival probability \cite[Sec.~IV]{SinopoliEtAl2004}.
When~\eqref{eq:critical} fails, the age chain is still positive recurrent, but $\pi(c)=\infty$, and the long-run average cost is infinite.

\subsection{Two-Dimensional System}

\begin{figure}[t]
\centering
\includegraphics[width=0.9\columnwidth]{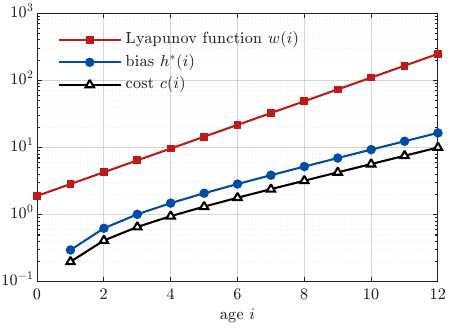}
\caption{Cost, canonical bias and Lyapunov function of the two-dimensional system for $\alpha=0.3$ and $\eta=1.5$. The cost and the bias vanish at the age $0$.}
\label{fig:certificate}
\end{figure}

Let
\begin{equation*}
  A=\begin{bmatrix}1.15&0.10\\0&0.90\end{bmatrix},\quad
  W=0.1\,I,\quad H=I,\quad\alpha=0.3.
\end{equation*}
Then $\alpha\,\sr(A)^2=0.39675<1$.
The solution of~\eqref{eq:discounted_lyapunov} is
\begin{equation*}
  X_\alpha=\begin{bmatrix}0.16702&0.00517\\0.00517&0.13210\end{bmatrix},
\end{equation*}
and~\eqref{eq:ncs_average_cost} gives $\rho^*=0.08974$.
The bias follows from~\eqref{eq:ncs_F_closed} and~\eqref{eq:ncs_bias_general}.
For the Lyapunov function we choose $\eta=1.5$.
The solution of~\eqref{eq:S_lyapunov} is
\begin{equation*}
  Y=\begin{bmatrix}8.45070&2.08996\\2.08996&2.84159\end{bmatrix},
\end{equation*}
which gives $\kappa_H=0.46544$, $\overline C=1.05117$ and $w(i)=1.91122\cdot1.5^{\,i}$.
Table~\ref{tab:ncs_certificate} lists the first values.
The conditions hold at every state and not only on the displayed range, because $(\taboo w)(i)=0.45\,w(i)$ and the killing probability is $0.55$ at every $i$.
Fig.~\ref{fig:certificate} shows the three functions on a logarithmic scale.
The cost and the bias grow at the rate $\sr(A)^2=1.3225$ and the Lyapunov function grows at the rate $\eta=1.5$.
The analytical bound in Proposition~\ref{prop:ncs} guarantees
$|h^*|\le w$ at every age.

\begin{table}[t]
\centering
\caption{Certificate Values for the Two-Dimensional System}
\label{tab:ncs_certificate}
\begin{tabular}{c|ccc}
\hline
$i$ & $c(i)$ & $h^*(i)$ & $w(i)$\\
\hline
0 & 0 & 0 & 1.9112\\
1 & 0.2000 & 0.2991 & 2.8668\\
2 & 0.4143 & 0.6295 & 4.3003\\
3 & 0.6590 & 1.0167 & 6.4504\\
4 & 0.9534 & 1.4914 & 9.6756\\
\hline
\end{tabular}
\end{table}

The bounds of Proposition~\ref{prop:inexact} are $\underline\rho_4=0.08828$, $\underline\rho_8=0.08970$ and $\overline\rho=0.28668$.
The lower bound with $n=8$ agrees with the gain to three digits.
The upper bound is about $3.2$ times the gain.
Both are obtained without the solution of~\eqref{eq:discounted_lyapunov}.

By Definition~\ref{def:twist}, $K(i,i+1)=\alpha\,w(i+1)/w(i)=0.45$ and $K(i,\partial)=0.55$ at every $i$.
The transformed chain is absorbed at each step with the same probability, and hence its absorption time is geometric and finite almost surely.
This is the content of~\ref{itm:l5} to~\ref{itm:l7} for the constant supermartingale and variant.

\begin{figure}[t]
\centering
\includegraphics[width=0.9\columnwidth]{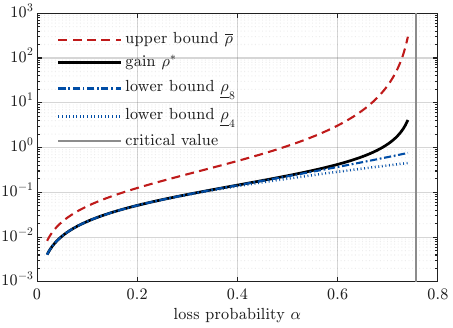}
\caption{Gain and certified bounds of Proposition~\ref{prop:inexact} as functions of the loss probability. The rate $\eta$ is the midpoint of the interval in~\eqref{eq:eta_choice}.}
\label{fig:gain}
\end{figure}

Table~\ref{tab:ncs_sweep} repeats the computation for four loss probabilities.
The critical loss probability is $1/\sr(A)^2=0.75614$, which corresponds to the arrival probability $0.24386$.
For each $\alpha$ the rate $\eta$ is the midpoint of the interval in~\eqref{eq:eta_choice}.
For $\alpha=0.3$ the midpoint is $\eta=2.32792$ and differs from the rate $\eta=1.5$ of Table~\ref{tab:ncs_certificate}.
Both rates are admissible, and the midpoint gives the upper bound $0.25446$ in place of $0.28668$.
The gain grows without bound as $\alpha$ approaches the critical value.
The killing probability $1-\alpha\eta$ tends to zero at the same time, because the interval in~\eqref{eq:eta_choice} closes.
The certificate exists for every loss probability below the critical value, and its margin measures the distance to that value.
Fig.~\ref{fig:gain} shows the gain and the certified bounds over the whole range of loss probabilities.
The lower bounds are tight for small $\alpha$ and lose accuracy near the critical value, where long ages carry most of the stationary cost.
The upper bound stays within one order of magnitude of the gain up to $\alpha=0.6$.

\begin{table}[!t]
\centering
\caption{Gain and Certificate for Four Loss Probabilities}
\label{tab:ncs_sweep}
\begin{tabular}{c|ccccc}
\hline
$\alpha$ & $\rho^*$ & $\eta$ & $1-\alpha\eta$ & $w(0)$ & $\overline\rho$\\
\hline
0.1 & 0.02242 & 5.66125 & 0.43388 & 0.1052 & 0.04904\\
0.3 & 0.08974 & 2.32792 & 0.30162 & 0.6387 & 0.25446\\
0.5 & 0.23554 & 1.66125 & 0.16938 & 3.2912 & 1.08814\\
0.7 & 1.19982 & 1.37554 & 0.03713 & 93.8262 & 24.66456\\
\hline
\end{tabular}
\end{table}

\begin{figure}[t]
\centering
\includegraphics[width=0.9\columnwidth]{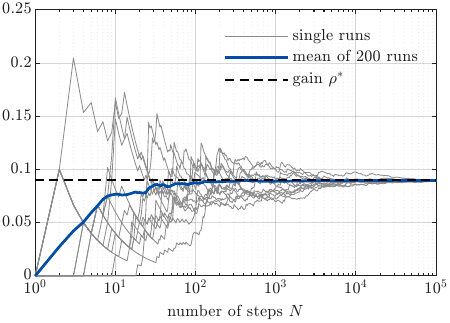}
\caption{Running average of the cost along simulated runs of the age chain for $\alpha=0.3$. Twelve of the $200$ runs, selected randomly, are shown.}
\label{fig:average}
\end{figure}

Fig.~\ref{fig:average} compares the gain with simulated runs of the age chain for $\alpha=0.3$.
Every run starts at the age $0$ and has $10^5$ steps.
The running averages of the single runs converge to the gain, and the mean over $200$ runs is $0.08972$ at the last step.

\section{Conclusion}
\label{sec:conclusion}

We provided two rules for lower and upper bounds on the long-run average cost of a countable Markov chain.
The Lyapunov rule replaces the growth condition of the classical verification theorem by a superharmonic Lyapunov function for the killed chain whose harmonic part is zero at the reference state, and it is complete on irreducible positive recurrent chains with one canonical bias.
The local rule replaces that last limit by a reachability certificate for the transformed chain, and every one of its conditions is an inequality over one transition.
Both rules are sound on every countable chain and complete for irreducible positive recurrent chains whose expected absolute cost in steady state is finite, and the cost need not be bounded.
For the networked control system the local certificate follows from two Lyapunov equations, and a certificate for a finite upper bound exists exactly above the critical arrival probability. Future work includes the extension of the rules to controlled chains and the average-cost optimality inequality, the extension to general state spaces, and algorithms that compute certificates within given function classes.

\appendices

\section{Proof of Lemma~\ref{lem:hitting}}

\begin{proof}
Let $E=\E_z\bigl[\sum_{t=0}^{\retz-1}f(X_t)\bigr]$.
By~\eqref{eq:kac}, $E=\pi(f)/\pi(z)<\infty$.
For $x=z$ the claim holds because $\hitz=0$.
Fix $x\ne z$ and let $A=\{\tau_x<\retz\}$ be the event that the chain started at $z$ reaches $x$ before it returns to $z$.

We show $\Prob_z(A)>0$.
By irreducibility there are states $z=y_0,y_1,\dots,y_k=x$ with $p(y_{i+1}\mid y_i)>0$ for every $i<k$.
Let $j$ be the largest index with $y_j=z$.
The chain started at $z$ follows $y_j,y_{j+1},\dots,y_k$ with positive probability, and on that event it reaches $x$ at time $k-j$ without a return to $z$.

On $A$ the segment $X_{\tau_x},\dots,X_{\retz-1}$ starts at $x$ and ends just before the first visit to $z$.
Since $f\ge0$,
\begin{equation*}
  E\ge\E_z\Bigl[\ind_A\sum_{t=\tau_x}^{\retz-1}f(X_t)\Bigr]
  =\Prob_z(A)\,\E_x\Bigl[\sum_{t=0}^{\hitz-1}f(X_t)\Bigr],
\end{equation*}
where the equality is the strong Markov property at $\tau_x$ and holds in $[0,\infty]$.
Division by $\Prob_z(A)$ proves the claim.
\end{proof}

\section{Proof of Lemma~\ref{lem:cesaro}}

\begin{proof}
The expected values $\pi(c^+)$ and $\pi(c^-)$ in steady state are finite.
It therefore suffices to prove the claim for $c\ge0$ with $\pi(c)<\infty$.
Let $T_0=\hitz$ and let $T_0<T_1<T_2<\cdots$ be the successive visits to $z$.
For $i\ge1$ let $L_i=T_i-T_{i-1}$ be the length and $C_i=\sum_{T_{i-1}\le t<T_i}c(X_t)$ the cost of the $i$-th cycle.
By the strong Markov property the pairs $(L_i,C_i)$ are independent and identically distributed, and by~\eqref{eq:kac} their means are $\E[L_1]=1/\pi(z)$ and $\E[C_1]=\pi(c)/\pi(z)$.
Let $K_N$ be the number of indices $i\ge1$ with $T_i\le N$, and let $D=\sum_{t<T_0}c(X_t)$.
Lemma~\ref{lem:hitting} gives $\E_x[D]<\infty$.
Since $c\ge0$,
\begin{equation}
\label{eq:sandwich}
  \sum_{i=1}^{K_N}C_i\le\sum_{t=0}^{N-1}c(X_t)\le D+\sum_{i=1}^{K_N+1}C_i.
\end{equation}
The index $K_N+1$ is a stopping time for the filtration generated by $T_0$ and the cycles, and $\E[K_N+1]\le N+1$.
Wald's identity \cite[Ch.~V]{Asmussen03} gives $\E\bigl[\sum_{i\le K_N+1}C_i\bigr]=\E[K_N+1]\,\E[C_1]$, and the elementary renewal theorem in its delayed form \cite[Ch.~V]{Asmussen03} gives $\E[K_N]/N\to\pi(z)$.
The right inequality in~\eqref{eq:sandwich} therefore gives
\begin{equation*}
  \limsup_{N\to\infty}\frac1N\,\E_x\Bigl[\sum_{t<N}c(X_t)\Bigr]\le\pi(z)\,\E[C_1]=\pi(c).
\end{equation*}
For the lower bound, $\E\bigl[\sum_{i\le K_N}C_i\bigr]=\E[K_N+1]\,\E[C_1]-\E[C_{K_N+1}]$ and $\E[C_{K_N+1}]\le\E[\max_{i\le N+1}C_i]$.
For every $n\ge1$ and $a>0$,
\begin{equation*}
  \E\bigl[\max_{i\le n}C_i\bigr]\le a+n\,\E\bigl[C_1\ind_{\{C_1>a\}}\bigr].
\end{equation*}
With $a=\varepsilon n$, division by $n$ and the limits $n\to\infty$ and then $\varepsilon\to0$ show that the left side grows slower than $n$.
The left inequality in~\eqref{eq:sandwich} then gives $\liminf_N\frac1N\E_x\bigl[\sum_{t<N}c(X_t)\bigr]\ge\pi(c)$.
\end{proof}

\section{Derivation of the Decomposition~\eqref{eq:riesz}}
\label{app:riesz}

Let $w:\States\to[0,\infty)$ be superharmonic for $\taboo$ and set $g=w-\taboo w\ge0$.
For every $N\ge1$ the sum telescopes,
\begin{equation*}
  \sum_{n=0}^{N-1}\taboo^{\,n}g
  =
  \sum_{n=0}^{N-1}
  \bigl(\taboo^{\,n}w-\taboo^{\,n+1}w\bigr)
  =
  w-\taboo^{\,N}w.
\end{equation*}
Letting $N\to\infty$ gives~\eqref{eq:riesz}.
Moreover, $0\le\taboo^{\,n}w\le w$ and $(\taboo w)(x)<\infty$. Hence,
dominated convergence gives
\begin{equation*}
  (\taboo w_\infty)(x)
  =
  \lim_{n\to\infty}(\taboo^{\,n+1}w)(x)
  =
  w_\infty(x).
\end{equation*}
Thus, $w_\infty$ is harmonic.
\section{Proof of Theorem~\ref{thm:expected}}
\label{app:expected}

\begin{proof}
We prove the first statement.
Fix $x$ and $t$.
By~\ref{itm:c2} the variable $h(X_{t+1})$ is integrable. Hence, the Markov property gives $\E_x[h(X_{t+1})\mid\mathcal F_t]=(Ph)(X_t)$ almost surely, and the conditional Jensen inequality gives $\E_x|(Ph)(X_t)|\le\E_x|h(X_{t+1})|<\infty$.
By~\ref{itm:c1}, $c(X_t)\ge\rho+h(X_t)-(Ph)(X_t)$.
Hence
\begin{equation*}
  \E_x[c(X_t)^-]\le|\rho|+\E_x|h(X_t)|+\E_x|h(X_{t+1})|<\infty.
\end{equation*}
The expectation of $\sum_{t<N}c(X_t)$ is therefore well defined in $(-\infty,\infty]$ and equals $\sum_{t<N}\E_x[c(X_t)]$.
Taking expectations in~\ref{itm:c1} at the state $X_t$ and using $\E_x[(Ph)(X_t)]=\E_x[h(X_{t+1})]$ from above
gives
\begin{equation*}
  \rho+\E_x[h(X_t)]\le\E_x[c(X_t)]+\E_x[h(X_{t+1})].
\end{equation*}
We sum this inequality over $t=0,\dots,N-1$.
The terms $\E_x[h(X_t)]$ with $1\le t\le N-1$ are finite and appear on both sides. Hence,~\eqref{eq:telescope} holds.
Dividing~\eqref{eq:telescope} by $N$ and rearranging gives
\begin{equation*}
  \frac1N\E_x\Bigl[\sum_{t=0}^{N-1}c(X_t)\Bigr]
  \ge
  \rho+\frac{h(x)}{N}
  -\frac1N\E_x[h(X_N)].
\end{equation*}
The limit inferior of the left side is $J(x)$. On the right side, $h(x)/N\to0$, and the limit inferior of $-\frac1N\E_x[h(X_N)]$ equals $-\limsup_{N\to\infty}\frac1N\E_x[h(X_N)]$, which is nonnegative by~\ref{itm:c2}. Hence $J(x)\ge\rho$. The second statement follows in the same way with all inequalities reversed.
\end{proof}

\section{Proof of Lemma~\ref{lem:poisson}}
\label{app:poisson}

\begin{proof}
Let $x\ne z$.
The chain started at $x$ has $\hitz\ge1$, and the sum in Definition~\ref{def:canonical} starts with the term $c(x)-\rho^*$.
By the Markov property at time $1$, the rest of the sum has the conditional expectation $h^*(X_1)$, where $h^*(X_1)=0$ on the event $X_1=z$.
Hence $h^*(x)=c(x)-\rho^*+(Ph^*)(x)$.
Let $x=z$.
The same decomposition of the first cycle gives
\begin{align*}
  c(z)-\rho^*+(Ph^*)(z)
  &=\E_z\Bigl[\sum_{t=0}^{\retz-1}\bigl(c(X_t)-\rho^*\bigr)\Bigr]\\
  &=\frac{\pi(c)-\rho^*}{\pi(z)}=0=h^*(z)
\end{align*}
by~\eqref{eq:kac}, applied to $c^+$, $c^-$ and the constant function $1$.
The exchanges of expectation and summation are justified by $G<\infty$ and by~\eqref{eq:kac} with $f=c'$.
\end{proof}

\section{Proof of Theorem~\ref{thm:reduction}}
\label{app:reduction}

\begin{proof}
By~\ref{itm:f3}, it suffices to show that $\E_x[w(X_t)]<\infty$ for every $t\in \mathbb N$ and $x\in\States$, and  $\frac1t\E_x[w(X_t)]\to0$ as $t\to\infty$, for every $x\in\States$. Set $a_n=(\taboo^{\,n}w)(z)$. By~\ref{itm:f2} and monotonicity of $\taboo$, $\taboo^{\,n+1}w\le\taboo^{\,n}w\le w$. Hence $0\le a_n\le w(z)$. Fix $x$ and $t\ge1$. Splitting trajectories according to whether
they avoid $z$ during times $1,\ldots,t$, and otherwise according to
their last visit to $z$ up to time $t$, the Markov property and
\eqref{eq:taboo-iter} give the last-exit bound~\eqref{eq:lastexit}, in which the sum equals $\sum_{n=0}^{t-1}a_n$.
Since $\taboo^{\,t}w\le w$ and $a_n\le w(z)$,
\begin{equation*}
  \E_x[w(X_t)]
  \le
  w(x)+t\,w(z)
  <\infty.
\end{equation*}

By~\ref{itm:f4}, $a_n\to0$. Therefore, the averages
$\frac1t\sum_{n=0}^{t-1}a_n$ also converge to zero.  Dividing~\eqref{eq:lastexit} by $t$ and using
$(\taboo^{\,t}w)(x)\le w(x)$ gives
\begin{equation*}
  \frac1t\E_x[w(X_t)]
  \le
  \frac{w(x)}t
  +
  \frac1t\sum_{n=0}^{t-1}a_n
  \longrightarrow0.
\end{equation*}
Finally, \ref{itm:f3} yields
$
  \E_x[|h(X_t)|]\le\E_x[w(X_t)]<\infty
$
and
$
  \frac{|\E_x[h(X_t)]|}{t}
  \le
  \frac{\E_x[w(X_t)]}{t}
  \longrightarrow0.
$
Thus the growth condition holds in both forms, and the result follows from
Theorem~\ref{thm:expected}.
\end{proof}

\section{Proof of Theorem~\ref{thm:reach}}
\label{app:reach}

\begin{proof}
Let $(Z_n)_{n\in\N}$ be the chain $M$ started at $s_0$.

\emph{Soundness.}
Let $\Inv$, $V$ and $U$ be a reachability certificate at $s_0$.
By~\ref{itm:r1} the chain stays in $\Inv$.
A state $s\ne\partial$ of $\Inv$ has $U(s)\ge1$, because~\ref{itm:r3} requires a successor with a smaller value of $U$.
By~\ref{itm:r2} the process $V(Z_n)$ is a nonnegative supermartingale.
It converges almost surely to a finite limit \cite[Sec.~35]{Billingsley95}, and hence $\sup_nV(Z_n)<\infty$ almost surely.
Fix $r\in\N$, let $D_r=\{s\in\Inv \mid V(s)\le r\}$, and let $\nu_r\in\N$, with $\nu_r\ge1$, be a bound of $U$ on $D_r$, which exists by~\ref{itm:r3}.
By~\ref{itm:r3}, from a state $s\ne\partial$ of $D_r$ the chain moves with probability at least $\varepsilon_r$ to a state with a smaller value of $U$.
Hence, if the chain is in $D_r\setminus\{\partial\}$ at a time $n$, then with conditional probability at least $\varepsilon_r^{\nu_r}$ it leaves $D_r$ or reaches $\partial$ during the next $\nu_r$ steps, because $\nu_r$ consecutive decreases inside $D_r$ from a value at most $\nu_r$ lead to the value $U=0$, i.e., to $\partial$.
We apply this bound at the times $0,\nu_r,2\nu_r,\dots$ and use the Markov property.
The probability that the chain stays in $D_r\setminus\{\partial\}$ up to the time $k\nu_r$ is at most $(1-\varepsilon_r^{\nu_r})^{k}$, which tends to zero.
Hence, almost surely, the chain leaves $D_r$ or reaches $\partial$.
On the event $\sup_nV(Z_n)\le r$ the chain never leaves $D_r$, and therefore it reaches $\partial$ almost surely on this event.
The union of these events over $r\in\N$ has probability one.

\emph{Completeness.}
Suppose that the chain reaches $\partial$ from $s_0$ almost surely. If $s_0=\partial$, take $\Inv=\{\partial\}$,
$V=U=0$, and $\varepsilon_r=1$ for every $r\ge0$.
Henceforth assume $s_0\ne\partial$. Let $\Inv$ be the set of the states that are accessible from $s_0$, together with $\partial$.
This set satisfies~\ref{itm:r1}.
Every $s\in\Inv$ is visited from $s_0$ with positive probability, and hence the strong Markov property shows that the chain started at $s$ reaches $\partial$ almost surely.

Let $s_1,s_2,\dots$ enumerate $\Inv\setminus\{\partial\}$, with repetitions if this set is finite, and let $\Gamma_n=\{s_1,\dots,s_n\}$.
Let $g_n(s)$ be the probability that the chain started at $s$ visits a state outside $\Gamma_n\cup\{\partial\}$ before it reaches $\partial$.
Then $g_n(\partial)=0$, and $g_n(s)=1$ for $s\notin \Gamma_n\cup\{\partial\}$.
For $s\in\Gamma_n$ the Markov property gives $g_n(s)=\sum_{s'}P_M(s,s')g_n(s')$.
Since $g_n\le1$, the function $g_n$ satisfies the inequality in~\ref{itm:r2} at every state of $\Inv$.
The chain started at $s$ visits finitely many states before it reaches $\partial$, and hence $g_n(s)\to0$ as $n\to\infty$ for every $s$.
We choose indices $n_1<n_2<\cdots$ such that $g_{n_k}(s_j)\le2^{-k}$ for every $j\le k$, and we set
\begin{equation*}
  V=\ind_{\Inv\setminus\{\partial\}}+\sum_{k=1}^{\infty}g_{n_k}.
\end{equation*}
At the state $s_j$ all terms with $k\ge j$ are at most $2^{-k}$, and hence $V$ is finite.
The indicator satisfies the inequality in~\ref{itm:r2}, and the sum satisfies it by Tonelli's theorem.
Moreover $V(\partial)=0$ and $V\ge1$ on $\Inv\setminus\{\partial\}$.
This is~\ref{itm:r2}.
If $V(s)\le r$ for a state $s\ne\partial$, then $s$ lies outside at most $r$ of the increasing sets $\Gamma_{n_k}$, and therefore $s\in\Gamma_{n_{\lfloor r\rfloor+1}}$.
Hence every sublevel set of $V$ is finite.
Let $U(s)$ be the length of the shortest path of positive probability from $s$ to $\partial$.
It is finite on $\Inv$, and $U(\partial)=0$.
Every state $s\ne\partial$ of $\Inv$ has a successor $s'$ with $U(s')=U(s)-1$.
On a finite sublevel set of $V$, $U$ is bounded, and the probabilities of these transitions have a positive minimum $\varepsilon_r$.
If the sublevel set contains no state other than $\partial$, set $\varepsilon_r=1$.
This is~\ref{itm:r3}.
\end{proof}

\bibliographystyle{IEEEtran}
\bibliography{certificates}

\end{document}